\documentclass[fleqn,usenatbib]{mnras}

\usepackage{newtxtext,newtxmath}

\usepackage[T1]{fontenc}

\DeclareRobustCommand{\VAN}[3]{#2}
\let\VANthebibliography\thebibliography
\def\thebibliography{\DeclareRobustCommand{\VAN}[3]{##3}\VANthebibliography}

\usepackage{graphicx}	
\usepackage{amsmath}	
\usepackage{pdflscape}
\usepackage{xspace}
\usepackage{stfloats}

\newcommand{\Sec}[1]{{\protect\hyperref[sec:#1]{Section~\ref*{sec:#1}}}}

\newcommand{\Fig}[1]{{\protect\hyperref[fig:#1]{Figure~\ref*{fig:#1}}}}

\newcommand{\subFig}[2]{{\protect\hyperref[fig:#1]{Figure~\ref*{fig:#1}~#2}}}
\newcommand{\Equ}[1]{{\protect\hyperref[equ:#1]{Equation~\ref*{equ:#1}}}}

\newcommand{\Tab}[1]{{\protect\hyperref[tab:#1]{Table~\ref*{tab:#1}}}}

\newcommand{\App}[1]{{\protect\hyperref[app:#1]{Appendix~\ref*{app:#1}}}}

\newcommand{\AP}{{A{\scriptsize STRO}P{\scriptsize HOT}}\xspace}
\newcommand{\PT}{{\scriptsize PYTORCH}\xspace}

\date{XXXX; YYYY; ZZZZ}
\pubyear{2023}

\title[AstroPhot]{AstroPhot: Fitting Everything Everywhere All at Once in Astronomical Images}

\author[C. Stone et al.]{
\href{https://orcid.org/0000-0002-9086-6398}{Connor J. Stone},$^{1,2,3}$\thanks{connorstone628@gmail.com}
\href{https://orcid.org/0000-0002-8597-6277}{St{\'e}phane Courteau},$^{4}$
\href{https://orcid.org/0000-0002-3263-8645}{Jean-Charles Cuillandre},$^{5}$
\newauthor
\href{https://orcid.org/0000-0002-8669-5733}{Yashar Hezaveh},$^{1,2,3,6}$
\href{https://orcid.org/0000-0003-3544-3939}{Laurence Perreault-Levasseur}$^{1,2,3,6}$ and
\href{https://orcid.org/0000-0002-3929-9316}{Nikhil Arora}$^{7,8}$
\\
$^{1}$ Department of Physics, Universit{\'e} de Montr{\'e}al, Montr{\'e}al, Qu{\'e}bec, Canada\\
$^{2}$ Mila - Qu{\'e}bec Artificial Intelligence Institute, Montr{\'e}al, Qu{\'e}bec, Canada\\
$^{3}$ Ciela - Montr{\'e}al Institute for Astrophysical Data Analysis and Machine Learning, Montr{\'e}al, Qu{\'e}bec, Canada\\
$^{4}$ Department of Physics, Engineering Physics \& Astronomy, Queen{'}s University, Kingston, Ontario, Canada\\ 
$^{5}$ AIM, CEA, CNRS, Université Paris-Saclay, Université de Paris, F-91191 Gif-sur-Yvette, France\\
$^{6}$ Center for Computational Astrophysics, Flatiron Institute, 162 5th Avenue, 10010, New York, NY, USA \\
$^{7}$ New York University Abu Dhabi, PO Box 129188, Abu Dhabi, United Arab Emirates\\
$^{8}$ Center for Astro, Particle and Planetary Physics (CAP$^3$), New York University Abu Dhabi
}

\begin{document}
\label{firstpage}
\pagerange{\pageref{firstpage}--\pageref{lastpage}}
\maketitle

\begin{abstract}
    We present \AP, a fast, powerful, and user-friendly Python based astronomical image photometry solver. 
    \AP incorporates automatic differentiation and GPU (or parallel CPU) acceleration, powered by the machine learning library \PT. 
    \emph{Everything}: \AP can fit models for sky, stars, galaxies, PSFs, and more in a principled $\chi^2$ forward optimization, recovering Bayesian posterior information and covariance of all parameters.
    \emph{Everywhere}: \AP can optimize forward models on CPU or GPU; across images that are large, multi-band, multi-epoch, rotated, dithered, and more.
    \emph{All at once}: The models are optimized together, thus handling overlapping objects and including the covariance between parameters (including PSF and galaxy parameters).
    A number of optimization algorithms are available including Levenberg-Marquardt, Gradient descent, and No-U-Turn MCMC sampling. 
    With an object-oriented user interface, \AP makes it easy to quickly extract detailed information from complex astronomical data for individual images or large survey programs.
    This paper outlines novel features of the \AP code and compares it to other popular astronomical image modeling software. 
    \AP is open-source, fully Python based, and freely accessible here: \break\url{https://github.com/Autostronomy/AstroPhot}
\end{abstract}


\section{Introduction}\label{sec:introduction}

Upcoming astronomical surveys will push the limits of storage and computational capacity~\citep{Gilmozzi2005,EUCLID2017,Ivezic2019,Akeson2019}. 
New data analysis methods are needed to tackle the wealth of well characterized, deep and wide-field photometry data.
Beyond the large scale of new astronomical observations, other challenges are also faced by modern astronomy.
For instance, new observatories have unique and non-trivial point-spread-functions (PSFs) that require careful modelling to extract the best science results~\citep{Perrin2014, Abraham2014, Rowe2015,Schmitz2020,Liaudat2023}.
The Low Surface Brightness (LSB) domain also pushes the limits of telescope and detector capabilities; methods (and understanding of their uncertainties) must be adapted to meet these new challenges and requirements~\citep{InfanteSainz2020}.
Indeed, modelling the diffuse inter-galactic and inter-cluster medium is an active field of research enabled by modern LSB techniques~\citep{Mihos2019}.
The depth of modern imaging also reveals a crowded Universe with many overlapping systems, even for objects traditionally viewed as isolated.
With blended sources near the confusion limit, analysis requires increasingly detailed models to fully take advantage of the available data.
Likewise, the multi-band and multiplexed nature of modern surveys require sufficiently powerful analysis tools to realize their full scientific potential. 
Finally, the shear scale of modern surveys necessitates fully automated, flexible, and user friendly pipelines to extract information on reasonable timescales.

Analysis based on $\chi^2$ 2D forward modelling is a powerful tool for extracting the available information embedded in astronomical images. 
In this framework, one may use models for the light distribution of stars and galaxies to analyse all the light sources in a given image.
These models are highly interpretable, providing total light, size, concentration, colour, and other properties of all the objects in the field. 
One may encode prior knowledge of the image by specifying an appropriate model to fit to the object (star, galaxy, edge-on galaxy, etc.).
However, 2D image modelling requires extensive computational resources as models must typically be evaluated many times over each pixel while iteratively converging to a solution.
To overcome the computational requirements, some simplifications can be made depending on the necessary information for a given science goal; for example one may simply add the flux from all pixels associated with an object to get an approximate total magnitude.
Alternatively, machine learning (ML) techniques can quickly solve complex inverse problems in a wide range of domains including image processing~\citep{Jia2020, Smith2021, Walmsley2022, Remy2022, Adam2023}.
Moreover, ML provides flexibility in defining complex prior models for astronomical images~\citep{Smith2022,Adam2022} which defy analytical treatments.
Despite the computational requirements, 2D image modelling benefits from interpretability, flexibility, convergence, and statistical guarantees; it is still the method of choice for a broad range of scientific tasks. 
Ultimately, various methods with complementary strengths exist, each being ideally suited to different problems.

The need for versatile and reliable image analysis software has motivated the development of a wide range of publicly-available packages.  
A non-exhaustive list includes: 
{\scriptsize XVISTA}~\citep{Lauer1985},
{\scriptsize ELLIPSE}~\citep{Tody1986},
{\scriptsize GIM2D}~\citep{Simard2002},
{\scriptsize GALFIT}~\citep{Peng2010},
{\scriptsize PYMORPH}~\citep{Vikram2010},
{\scriptsize GALAPAGOS}~\citep{Barden2012},
{\scriptsize MEGAMORPH}~\citep{Hausler2013},
{\scriptsize IMFIT}~\citep{Erwin2015}, 
{\scriptsize DISKFIT}~\citep{Sellwood2015}, 
{\scriptsize ISOFIT}~\citep{Ciambur2015}, 
{\scriptsize PHOTUTILS}~\citep{photutils},  
{\scriptsize GALSIM}~\citep{Rowe2015}, 
{\scriptsize THE TRACTOR}~\citep{Lang2016},
{\scriptsize PROFIT}~\citep{Robotham2017},
{\scriptsize GALIGHT}~\citep{Ding2020},
{\scriptsize SOURCEEXTRACTOR++}~\citep{Bertin2020},
{\scriptsize LENSTRONOMY}~\citep{Birrer2021},
{\scriptsize HOSTPHOT}~\citep{Muller2022},
{\scriptsize AUTOPHOT}~\citep{Brennan2022},
{\scriptsize PYAUTOGALAXY}~\citep{Nightingale2023},
and {\scriptsize MORPHOFIT}~\citep{Tortorelli2023}.
Many of these codes are interdependent and add new features to previous versions.
However, none of them employ GPU acceleration or automatic differentiation~\citep{Gunes2015}. 
Further, almost all are either written in a compiled language (typically C or FORTRAN) or act as wrappers to compiled core functionality making development and maintenance challenging.
The exceptions are {\scriptsize BANG}~\citep{Rigamonti2023} and {\scriptsize PYSERSIC}~\citep{Pasha2023} which do employ GPU acceleration. 
However these are specialized for S{\'e}rsic modelling with sampling methods (No-U-Turn and Nested sampling respectively) which are orders of magnitude slower than \AP's Levenberg-Marquardt (LM) based optimization (or equivalent to \AP's score based sampling methods).
Indeed, {\scriptsize PYTORCH} is already known to be a useful engine for scientific computing~\citep{Nikolic2018}.
\AP represents a significant effort to build a state-of-the-art photometry code with these powerful numerical tools and software libraries.

To address the above (and other) science challenges, fully utilize the latest computational resources, accelerate code development, and overall achieve state-of-the-art performance on modern datasets, we have developed \AP.
This code borrows numerical optimization and parallelization from the \PT Python package developed for ML.
\PT is used to accelerate classical $\chi^2$ based 2D image modelling by taking advantage of GPU acceleration and automatic differentiation.

This paper presents the \AP code and its capabilities.
It represents a substantial evolution over the original functionality of {A{\scriptsize UTO}P{\scriptsize ROF}}~\citep{Stone2021b}.
{A{\scriptsize UTO}P{\scriptsize ROF}} was developed to extract non-parametric 1D surface brightness profiles of single galaxy images using azimuthally-averaged isophotal ellipse fitting.
A key challenge in ellipse fitting is the extraction of global structures of the galaxy disk whilst ignoring spurious features such as overlapping foreground/background objects (e.g., stars and galaxies) and modeling non-axisymmetric features such as bars and spiral arms. 
To achieve a first order solution, {A{\scriptsize UTO}P{\scriptsize ROF}} used a regularization technique borrowed from ML to impose the prior of a smooth variation of the position angle and ellipticity across the surface of a galaxy.
{A{\scriptsize UTO}P{\scriptsize ROF}} could achieve robust results quickly without human intervention, however as a 1D ellipse fitting method, it could only return a smooth isophotal solution for a single object at a time.
This is unsuitable for a large variety of scientific use cases as galaxies are often in crowded environments (e.g., clusters) and pushing LSB limits rapidly escalates the overlapping issue.
This requires fully modelling all objects and accounting for PSF effects to disentangle sources of light.

To overcome this issue, the original {A{\scriptsize UTO}P{\scriptsize ROF}} was rebuilt into a full 2D forward modelling code: \AP.
This new code retains the ease of use and automatic initialization of the original version, whilst significantly improving other aspects of the tool such that it is now a fully general photometry code.
\AP can now fit all objects in an image (stars and galaxies, whether overlapping or isolated), account for complex PSF effects, jointly model across different photometric bands, and extract Bayesian posteriors. 
This paper describes the key updates added to the new code and demonstrates the capacity of \AP to tackle the most pressing challenges in astronomical image processing.
The new \AP continues to take advantage of ML methods, though this time in the form of efficient and powerful numerical libraries and/or techniques. 
These libraries make the \AP code ``differentiable'' such that derivatives can be propagated through the models giving access to powerful algorithms whose computation would otherwise be prohibitively expensive.
\AP can couple with existing ML frameworks and score-based methods, giving it unique capabilities amongst astronomical image modelling codes.

\Sec{howitworks} describes new available features and the Python code architecture, while \Sec{examples} highlights \AP's capabilities with a few illustrative cases.
Concluding remarks on the future of scientific computing and how \AP may expand the discovery space from modern deep astronomical imaging data are presented in \Sec{conclusions}.
Details on implementing \AP, the code repository, documentation, and up to date information can all be accessed at: \url{https://github.com/Autostronomy/AstroPhot}\footnote{The original version of {A{\scriptsize UTO}P{\scriptsize ROF}} \citep{Stone2021b} is available at: \url{https://github.com/Autostronomy/AutoProf}}.

\section{How \AP works}
\label{sec:howitworks}

This section gives an overview of the implementation and features available through \AP at a high level.
\AP is an open source project intended to meet numerous science goals simultaneously by ensuring a high degree of flexibility.
Some of these science goals require operational speeds far greater than possibly achieved with recent alternative tools (see \Sec{introduction}). 
Various science goals also call for specialized tools or domain specific modifications. 
\AP, being written in Python, makes it easy to adapt and expand the code to suit a wide variety of tasks.

\subsection{Image Modeling}
\label{sec:imagemodeling}

Cosmic photons are typically collected by a telescope and directed to a sensor where they are converted into a digital signal.
Image processing starts with an astronomical digital image\footnote{Astronomical images are often formatted as a \emph{FITS} file~\citep{Pence2010}, though other useful formats (e.g., HDF5) also exist.} for which a model of the celestial photon source(s) therein is desired. 
For \AP (and most similar packages), one must provide the image pixel scale, and optionally, a photometric zero-point and the position of the image on the sky.
Further steps depend on the scientific goals of the analysis.

Astronomical image modeling is a technique used to analyze and interpret images captured by telescopes and other observational instruments. 
Central to this approach is the process of constructing composite models that represent the light distribution in an image by combining individual sub-models. 
Each sub-model characterizes a specific object under investigation.
In a hierarchical representation, sometimes multiple sub-models are used to represent different aspects or features of a single object, such as bars and spiral arms of galaxies.
The primary objective of astronomical image modeling is to achieve a comprehensive and accurate representation of the light in an image by carefully selecting and fitting these sub-models.
This involves iteratively adjusting the parameters of each sub-model to minimize the discrepancies between the composite model and the observed data. 
Consequently, this method allows one to disentangle the contributions of various components or sources of light.
The resulting parametric representation of an image is far more scientifically informative than the matrix of pixel flux values which represent an unprocessed image.

Since many astronomical image fitting tasks are iterative in nature, one may incrementally add complexity to a model and visualize results along the way.
\AP includes a number of built-in diagnostic plotting features which help visualize the model and target images, residuals, and light profiles during fitting.
Such diagnostics can dramatically speed up the time to convergence by helping to identify poor initial conditions, spurious image features, or mismatched units (to name a few basic examples).

\subsubsection{A Zoo of Models}
\label{sec:modelzoo}

A number of models are readily available in \AP. 
Users may also easily add other models with the object oriented Python interface.
\AP is equipped with most standard 2D model profiles such as the Exponential, S{\' e}rsic~\citep{sersic1963}, Gaussian, Nuker~\citep{Lauer1995}, and Moffat~\citep{Moffat1969} functions as well as models for flat or tilted sky planes.
\Fig{modelzoo} presents most of the currently available models in \AP.
More will be added in future updates.
All models can be superimposed to build arbitrarily sophisticated representations of an astronomical image.
Many of the models are common in astronomy, for example the ``sersic galaxy model'' is a typical S{\'e}rsic profile with elliptical isophotes.
However, some of the models are less common or even unique to \AP, and deserve further description.
Below we discuss the \emph{psf star model}, \emph{zernike star model}, \emph{fourier}, \emph{ray}, and \emph{warp} models as well as the \emph{spline} variants available. 

The ``psf star model'' is a fully general PSF pixel map which the user can provide.
In \Fig{modelzoo}, it is shown as an Airy disk but any image could be provided.
The pixel map is interpolated on the fly using bilinear interpolation, this is very fast but imprecise.
Thus one may provide the PSF at higher resolution, essentially to push the expensive high quality interpolation stage to a single operation outside the many optimization steps.

The ``zernike star model'' is an alternate star model which uses Zernike polynomials to decompose a star into modes of increasing complexity.
The user may select the order ``n'' to use all modes up to that level, or may individually select which modes to use (for example selecting only radial modes).
Zernike polynomials allow for very general PSF models to be fitted while giving the user control over the trade off between expressiveness of the model and computational complexity.

The column of ``fourier'' type models in \Fig{modelzoo} represent arbitrary perturbations from a pure ellipse as defined by a select number of frequencies.
This is done by adjusting the radius as a function of angle around the ellipse as follows:

\begin{equation}\label{equ:fouriermodeperturbations}
    \begin{aligned} 
        R(\theta) &= R_0(\theta)e^{\sum_m A_m\cos(m(\theta + \phi_m))},    
    \end{aligned}
\end{equation}

where $R$ is the radius after perturbation, $\theta$ is the angle relative to the major axis of the galaxy, $R_0$ is the original radius on the ellipse, $m$ is an integer mode for the perturbation (chosen before optimization), $A_m$ is the amplitude of the perturbation, and $\phi_m$ is the phase (direction) of the perturbation.
A mode of 1 displaces the center, 2 is nearly degenerate with ellipticity and should be avoided, 3 makes the model lopsided, 4 is similar to boxy/disky isophotes though more rounded, and higher orders involve sinusoidal adjustments of finer precision.
During fitting, the $A_m$ and $\phi_m$ parameters are optimized, the number of these coefficients is determined by the selected number of modes $m$.
Fourier models are ideal for representing irregular or perturbed systems, including complex features like streams and tidal tails in LSB enabled datasets.
However, the increased complexity of these models makes for a challenging implementation, achieving convergence on these models can be very slow or may fail entirely if the initial parameters are not set accurately.
For a powerful and simple application, the first mode can accurately detect if a system has been perturbed by determining the degree to which the center is offset relative to the outskirts.

The two columns of ``ray'' galaxy models represent ways to subdivide a model by allowing multiple radial profiles to be fit.
The first column has one model for the major axis and a different model for the minor axis.
Both models have the same profile form (exponential, S{\'e}rsic, etc.) but may have different values for the parameters of the model. 
The two models have a smooth transition using a cosine weighting scheme.
The second column is similar except that now four models are fit along the minor axis, major axis, and the two diagonals.
An arbitrary number of ``rays'' are allowed at the cost of additional parameters and fewer pixels to fit them each time.
A similar ``wedge'' model lacks a smooth transition, as it abruptly jumps from one model to another.
These ``ray'' and ``wedge'' models can be useful for examining axisymmetry in a galaxy by relaxing the constraint in a controlled way.

The final column of ``warp'' galaxy models allows for the position angle and axis ratio to vary as a function of radius.
The latter is nontrivial. 
In elliptical isophote fitting, this is done by requiring ellipses on the sky have the same brightness~\citep{Lauer1985, Tody1986, Courteau1996, Stone2021b}.
However, enforcing elliptical isophotes in a forward modelling context is challenging and such a configuration may not best represent a galaxy.
Inspired by \citet{Peng2010} (and similar parametric modeling packages), we extend the global ellipse transformation into a radially varying one as:

\begin{equation}\label{equ:spline}
\begin{aligned}    
    R_1 &= \sqrt{X_1^2 + Y_1^2} \\
    X_2 &= \cos(\theta(R_1))X_1 - \sin(\theta(R_1))Y_1 \\
    Y_2 &= \sin(\theta(R_1))X_1 + \cos(\theta(R_1))Y_1 \\
    R_2 &= \sqrt{X_2^2 + (Y_2/q(R_1))^2} \\
    I &= I(R_2),
\end{aligned}
\end{equation}

\noindent where $R_1, X_1, Y_1$ are the on-sky radius and coordinates, $R_2, X_2, Y_2$ are the transformed coordinates, $\theta(R_1)$ is the radially varying PA function, $q(R_1)$ is the radially varying axis ratio function, and $I(R_2)$ is the surface brightness profile in the transformed coordinates.
We add a further layer of realism by transforming $X_1, Y_1$ according to a global PA and axis ratio before applying the transformations in \Equ{spline}.
Thus the coordinates $X_1,Y_1$ can be roughly interpreted as the ``face-on'' coordinates of a given galaxy for small inclinations.
In principle there is a degeneracy between the two ellipse transformations.
However in practice, this does not pose a problem for an object with enough structure (see \Sec{examplespline}).
By setting constraints on the global $q$ and PA values one can express a physical interpretation on the results.
For example, a disk system may have freely varying global $q$ and PA while an elliptical system may have fixed values since an inclination is less meaningful in that scenario. 

Most of the rows in \Fig{modelzoo} represent common model types adopted in astronomy.
The spline model is more specialized to \AP and allows for a profile to be defined by a spline passing through a certain number of nodes.
The number and position of the nodes may be specified or \AP can choose its own.
Automatic node placement is done with logarithmic spacing such that higher resolution is available in the center and lower resolution in the outskirts.
The brightness associated with each node and the spline interpolation are represented in log space so that the model never evaluates negative fluxes.
There are two options to handle behavior after the last node: either the model fills extrapolated pixels with zeros, or it may perform linear (in log space, exponential in flux space) extrapolation using the last two nodes.
Spline models are extremely useful for analysis in the LSB regime where imposing a model may bias sensitive features.

The models described here can be combined to form a detailed representation of an astronomical image.
For survey style analysis or images with many objects, an external tool such as ProFound~\citep{Robotham2018} or MTObjects~\citep{teeninga2015} can segment an image and provide the basic object information to initialize \AP.
Further complications to the model would depend on the science goals such as modelling multiple images (\Sec{examplemultiband}), point sources (\Sec{examplepsfextract}), Bayesian uncertainties (\Sec{examplehmc}), and more.
Custom models may be required for specialized tasks; the models in \AP are organized in an inheritance structure that allows users to branch off and customize models without having to rewrite already integrated features such as inclination, superellipse, warping, etc.

\subsubsection{Pixel Integration}
\label{sec:pixelint}

By default, much like {\scriptsize PROFIT}~\citep{Robotham2017}, \AP attempts to ensure that pixels are sampled at least to an accuracy of 1 per\,cent.
In principle the 2D models in \AP represent pure infinite resolution models which are then projected onto pixels for comparison with an observed image.
This means that for a set of images with very high signal to noise the same model parameters should be recovered regardless of pixel size or dithering (position adjustments smaller than a pixel).
To achieve this kind of invariance, necessary for multi-band imaging, then requires integrating within pixels during model sampling.
If \AP were to simply take the center point of each pixel as a representation for the whole pixel, it would return incorrect fluxes often by over 20 (even 100) per cent.
This is because many widely used models such as the S{\'e}rsic display rapid shape variations, often within a single pixel.

For the sub pixel evaluation of a model, \AP must essentially solve many small integration problems simultaneously; this requires both an integration algorithm and an understanding of the accuracy of the integration.
A typical scheme for selecting pixels that require integration is to check for large derivatives.
However, any linear variation across a pixel can be modelled by the midpoint since it equals the average of a straight line, therefore selecting pixels based on the local derivative does not optimally allocate computational resources.
It is in fact large second order (and above) variations in the light across a pixel that cause the integration to fail.

\AP has several tools to efficiently handle sub-pixel integration.
At a high level, \AP performs two stages in its default configuration; a first pass evaluation of all pixels, and a second pass deeper integration of problematic pixels.
For the first pass, \AP uses the midpoint of every pixel since this is the fastest possible integration scheme and is accurate up to first order.
With this first pass complete, \AP uses this information to identify problematic regions of high second order derivatives.
In the second pass, problem pixels are evaluated using Gaussian quadrature, which efficiently determines the integral in a pixel for any function up to a given polynomial order\footnote{For Gaussian quadrature of order $n$, the resulting integral is accurate for polynomials of order $2n-1$ and requires $n^2$ evaluations within a pixel.}.
To begin, \AP uses third order quadrature meaning that nine evaluations within the pixels are used to accurately evaluate the integral if the light in a pixel can be represented by a fifth order polynomial.
However, some pixels (often only the centermost pixel) may still not have converged.
In this case, the remaining pixels are divided by a factor of five on each axis and quadrature of fifth order is performed (25 evaluations and accurate to 9th order polynomials).
Remaining sub-pixels are further divided by a factor of five (now at 25 times resolution) and quadrature of 7th order is performed (49 evaluations and accurate to 13th order polynomials).
At this point, \AP by default will exit as any further problem sub-sub-pixels are considered to have discontinuities (which cannot be resolved in this scheme).
Once this process is completed, most pixel fluxes which originally had poor accuracy, are now precise to below 1 per cent by several orders of magnitude.

Choosing which pixels are integrated is non-trivial due to a number of edge cases, \AP tries to reach a compromise that best preserves flux.
Selecting pixels with a high relative error is the most straightforward option, however for S{\'e}rsic indices less than 1 (such as a Gaussian at $n=0.5$), the higher order derivative values increase quickly with radius while the flux also decreases quickly.
This means that a relative error based selection method would spend a great deal of time integrating pixels whose flux may be many orders of magnitude lower than the noise level in the image or even below the numerical resolution of the computer.
Instead \AP uses a reference of the total image flux times a tolerance (by default 1 per cent) divided by the number of pixels, essentially creating a per-pixel error budget.
Every pixel must then converge to an error below $e = \frac{0.01F}{N}$ where $F$ is the total flux and $N$ is the number of pixels.
Therefore, faint pixels are allowed to have larger relative error, while bright pixels may need better than 1 per cent accuracy; over the whole image the error remains below 1 per cent.

The details of sub-pixel integration are complex and may change with time as \AP evolves in sophistication. 
If anything, further developments will maintain or exceed the current level of rigor in evaluation.
Further, users may already select from a number of options to suit their needs.
Sub-pixel integration may be turned off completely for improved speed, though this is not advised as the gains are marginal.
Alternatively, one may further push the tolerance for very high accuracy, and even apply Simpson's method integration at the first pass for orders of magnitude higher accuracy (at a cost of 4 times more evaluations).

\subsubsection{PSF convolution}

PSF convolution is performed using a PSF image, that can be provided by the user.
The PSF may be associated with an image in which case it will be shared by all models used to fit that image.
Alternatively, a PSF may be associated individually with a model, enabling fine grain control for example to build a PSF that varies across an image.
Alternatively, the PSF may itself be an \AP-built model, with its own parameters to be optimized during fitting.
The PSF may also have a higher resolution than the input image.
In this case, the model sampling and PSF convolution is performed at the higher resolution and finally integrated to compare with the observed image.
While most of these features are standard in astronomical image fitting codes (the exception being the live PSF models), they are substantial performance bottlenecks.
GPU acceleration largely bypasses this issue (\Sec{examplegpu}) as it is ideally suited for this kind of operation, providing order(s) of magnitude speedup.
Convolution can also be performed in ``direct'' mode which is efficient for small PSF images (especially when both PSF and model are small), or in ``FFT'' mode which is more efficient for large PSFs.

An important component of PSF convolution is the sub-pixel positioning of the model.
As an extreme case, consider a bright model with a radius smaller than one pixel (e.g., a star) and a PSF which is the size of one pixel, the position of the model within the pixel will determine how light is spread to neighboring pixels.
If it is exactly centered, then all the light should be in the one pixel, but if the model is near a corner of four pixels then the light should be evenly spread over the four pixels.
This is an extreme example, but the effect is still present for extended models and must be accounted for.
\AP operates under the assumption that the PSF was centered perfectly on a pixel, thus for proper convolution the model should also be centered on a pixel.
In the PSF convolution mode, \AP will enforce that the center of the model is located at the center of the nearest pixel before performing sampling (and sub-pixel integration) and then convolution.
Once this is completed, \AP uses interpolation to re-shift the model back to its true sub-pixel center position; the interpolation method can be selected as Lanczos or bilinear~\citep{Shannon1949, Burger2010}.
This process is very precise for most scenarios, however if an image is sampled more coarsely than the Nyquist limit then the sharp variations between pixels can make it impossible to accurately perform sub-pixel shifts.
In these cases one should provide a PSF at higher resolution such that modelling can be performed at or below the Nyquist limit.

\AP can also derive an optimized PSF (see \Sec{examplepsfextract}) before image modelling, or even during image modelling.
In principle one could fit a PSF model in an image with no stars, though that would only be accurate if the galaxy models were good representations of the true galaxies.
In general, the only reason stars are used to model PSFs is because the true underlying model is well-known (i.e., a delta function), hypothetically if the true model for a galaxy is known (i.e. it is well represented by a S{\'e}rsic) then there is no mathematical difference.
In fact, when a galaxy is the brightest object in an image it may have more constraining power for the PSF than any stars.
Since the convolution operation is symmetric between a galaxy image and the PSF image, there is no fundamental difference between them.
As well as it is possible to recover the galaxy properties while convolved with a fixed PSF, it is equally possible to recover PSF properties with a fixed galaxy model.
When neither model is fixed there are certain degenerate cases, though these degeneracies can easily be broken when the galaxy is inclined or there are multiple galaxies to fit.

\begin{figure*}
    \centering
    \includegraphics[width=\textwidth]{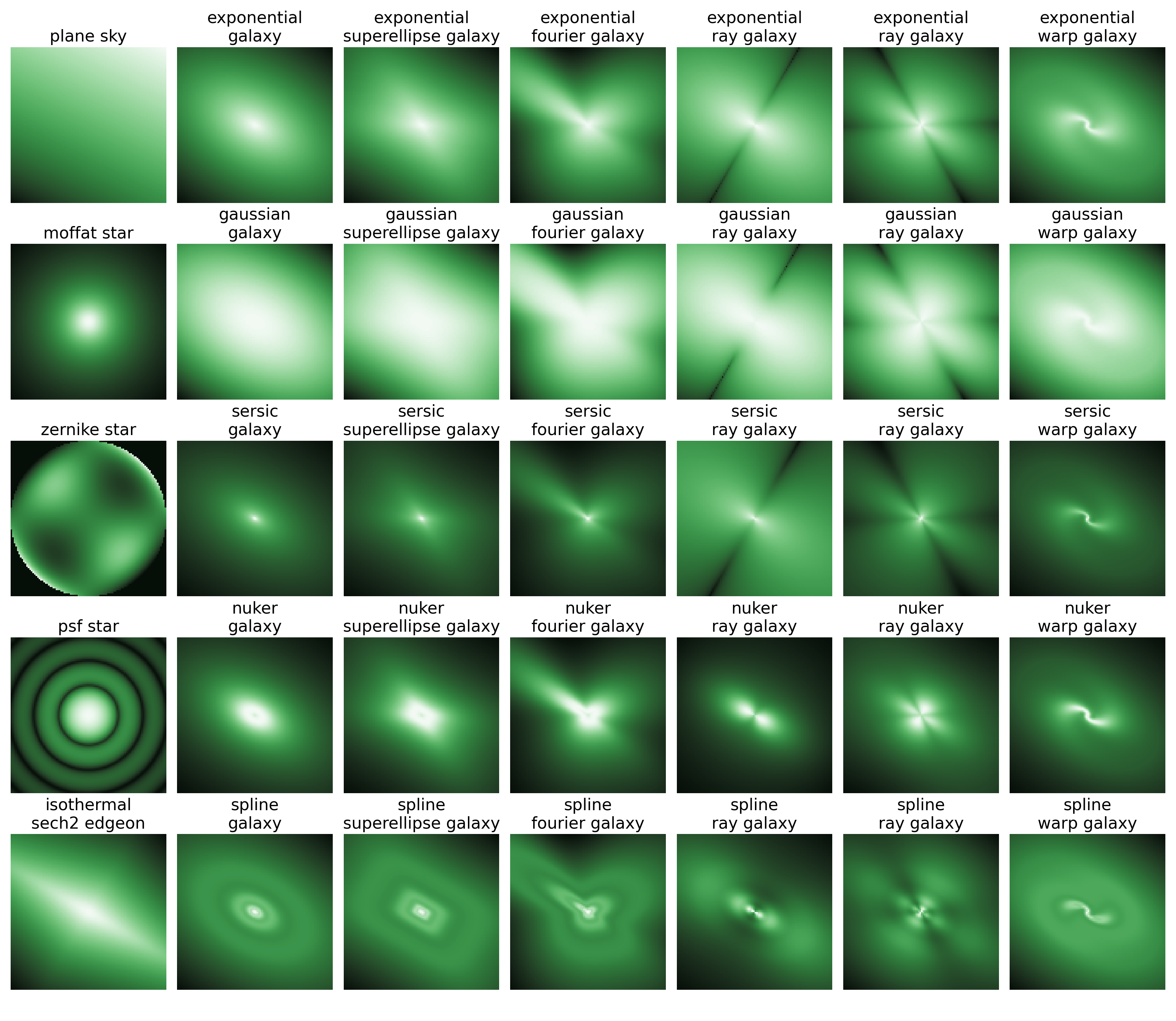}
    \caption{Currently available component models in \AP 
    (expected to grow over time).
    Models can represent sky properties, stars, galaxies, PSFs, and instrumental artefacts.
    Stacking these models enables the creation of an arbitrarily complex representation of an astronomical image.
    The name in each box is used by \AP to call an instance of the model (simply add ``model'' to the end).
    This figure was created using \AP's built-in plotting routines.
    A flowchart of currently available models in \AP is presented in \Fig{modelorgchart}. 
    }
    \label{fig:modelzoo}
\end{figure*}

\subsection{$\chi^2$ Optimization}

At its core, \AP is an image simulation software, it can create mock images specified by model parameters combined to produce arbitrarily complex representations of an observation.
In \AP, these models are differentiable, making it straightforward to build optimizers on the framework.
The most common kind of optimization is the maximum likelihood. 
Assuming pixels have plenty of signal, a reasonable representation is that observations (the actual values in the pixels) follow a Gaussian distribution; if the pixels are also independent, then their full likelihood is the product of likelihoods for each pixel:

\begin{equation}\label{equ:likelihood}
    P(D|\theta) = \Pi_i \frac{1}{\sqrt{2\pi\sigma_i^2}}e^{-\frac{(d_i - m(i,\theta))^2}{2\sigma_i^2}},
\end{equation}

\noindent where $D$ represents the full dataset, $\theta$ represents the model fitting parameters, $d_i$ is the $i$th observation (pixel), $m(i,\theta)$ is the model evaluated at the location of $d_i$, and $\sigma_i$ is the uncertainty on the observation $d_i$.
The index $i$ will range over all pixels in the image(s), where the relative weighting of the pixels is determined by the $\sigma_i$ values. 
The model parameters $\theta$ depend on what component is being used, and in fact users can easily add new models or modify existing models to frame a problem as they wish.
Note that maximizing the likelihood is equivalent to minimizing the negative log likelihood; this quantity is called the $\chi^2$ and can be written as:

\begin{equation}\label{equ:chi2}
    -\log(P(D|\theta)) = \chi^2 = \sum_i \frac{(d_i - m(i,\theta))^2}{2\sigma_i^2} + C.
\end{equation}

The model values $m(i,\theta)$ are the sum of fluxes from each component model at that pixel $i$ with its associated pixel size, photometric zeropoint, position, orientation, and other factors.
Since each model is associated with pixels of certain properties, the joint modeling of multiple images (e.g., multi-band modeling) also requires the creation of a separate component model for each image.
The user may then decide what parameters are shared between the multiple images.

For accurate modelling, \AP requires a variance ($\sigma^2$) estimate as input.
This variance image should have the same shape as the target image and contain the variance at each pixel.
In an ideal scenario where pixel fluxes have units of counts and all values are sufficiently large, then the variance map is simply equal to the image itself.
Typically, however, images are stacked, dithered, masked, flat fielded, and otherwise manipulated to make the individual pixel values more interpretable.
To this end, various packages, such as {\scriptsize SWARP}~\citep{Bertin2002} and {\scriptsize WEIGHTWATCHER}~\citep{Marmo2008}, provide a weight map as an output which can then be used by \AP.
An accurate variance map is important for both accurate fitted parameters and meaningful uncertainties.
In multi-band or multi-epoch fitting, it is even more critical as the variance map encodes the relative statistical power of the images.

\subsection{Implementation with \PT}
\label{sec:pytorch}

A core feature of \AP is its reliance on differentiable programming\footnote{Differentiable programming is a general paradigm in which derivatives may be propagated automatically through arbitrary code. It is widely used in Machine Learning, but also in other scientific disciplines.}
for its models, enabling the propagation of derivatives automatically through all calculations much faster than via finite differences techniques.
This is achieved with \PT\footnote{\PT is nearly a drop-in replacement for the widely used numpy~\citep{harris2020numpy} package in Python.} autograd, which constructs a computational graph tracking all mathematical operations when sampling a model.
With the computational graph, \PT can use the chain rule to work backwards from a $\chi^2$ value to a gradient in all input parameters.
Impressively, this backpropagation algorithm has the same computational complexity as sampling the model (typically only a factor of two longer) regardless of the number of parameters, while finite differences derivatives have a complexity equal to the forward model times the number of parameters.
Moreover, unlike finite differences, the autograd derivatives are exact up to numerical precision.

\PT includes a number of other beneficial properties for the sake of scientific computing.
Foremost, operations in \PT can be passed to a GPU\footnote{Specifically, any CUDA-enabled GPU.} for significant performance enhancement.
This is needed as models of image PSFs continue to grow in complexity and size~\citep{InfanteSainz2020}; a significant bottleneck for CPU based image analysis.
With GPU acceleration, it becomes possible to perform complete PSF convolution in reasonable timescales on forward models.
For instance, tests on \emph{A100} GPUs showed a $\sim$15x speedup relative to a \emph{AMD Milan 7413} CPU (see \Sec{examplegpu}). 
Even with a CPU, \PT takes advantage of all available computing resources and automatically parallelizes large image operations between cores.
However, the CPU-based speedup is only beneficial when a single image is being analyzed; in large projects with many parallel image fitting tasks, the CPU resources are already used at capacity.

By offloading the computational efficiency optimization to \PT, \AP could then be entirely written in Python.
Since there are no compiled functions in \AP\footnote{Aside from the \PT backend.}, the code is highly readable, flexible, and robust.
While we aim to make the documentation as complete as possible, even reading the code directly is quite straightforward for most operations.

\subsection{Automatic Initialization}
\label{sec:intialization}

\AP offers multiple optional modelling features.
For instance, a common challenge in image modelling is the determination of initial parameter values before applying the fitting algorithm.
To circumvent this issue, \AP includes automatic initialization for all standard models, based on simplified 1D fitting before proceeding with full 2D fitting. 
The initialization scheme is adapted from the initialization in \citet{Stone2021b}.
First \AP follows the light weighted center of mass in a small window until it finds a brightness peak.
Second, it uses the light weighted phase on circular isophotes to determine the position angle, it also tests a few discrete axis ratios to find a suitable starting value.
Third, a number of isophotes are sampled to give a 1D profile and the brightness parameters (e.g., $n, R_e, I_e$ for the S{\'e}rsic model) are optimized on this profile.
This places the profile in a reasonable starting location in the parameter space.
Each model may define its own initialization procedure. 
For example, a circularly symmetric star model will not require initializing the position angle or axis ratio; more complex models such as those with Fourier modes often simply initialize with zeros.
For a variety of situations, the manual initialization of parameter values before applying a fit is not needed; providing significant acceleration for interactive analysis.
For moderately crowded fields, one may model all the objects rather than simply attempting to mask them{\footnote{An avenue of further development is the possibility of initialization by a Neural Network, with the requirement of high performance in a large dynamic range of parameter space.}}.

Two further features aid in modelling many objects simultaneously.
\AP can use segmentation maps to build a large number of models accounting for every object in a field.
Identifying the appropriate model for each segment is still largely situation specific and is left to the user (though many segmentation mapping routines include flags to indicate if a segment is likely a star or galaxy).
However, adding dozens or hundreds of models in a large image can massively increase computation time if every model is evaluated at each pixel.
With this in mind, every model in \AP tracks its own ``window'' in the image; within which it will be evaluated.
One can therefore evaluate a model only in regions where it contributes light and save significant computational resources.

\begin{figure*}
    \centering
    \includegraphics[width = 0.3\textwidth]{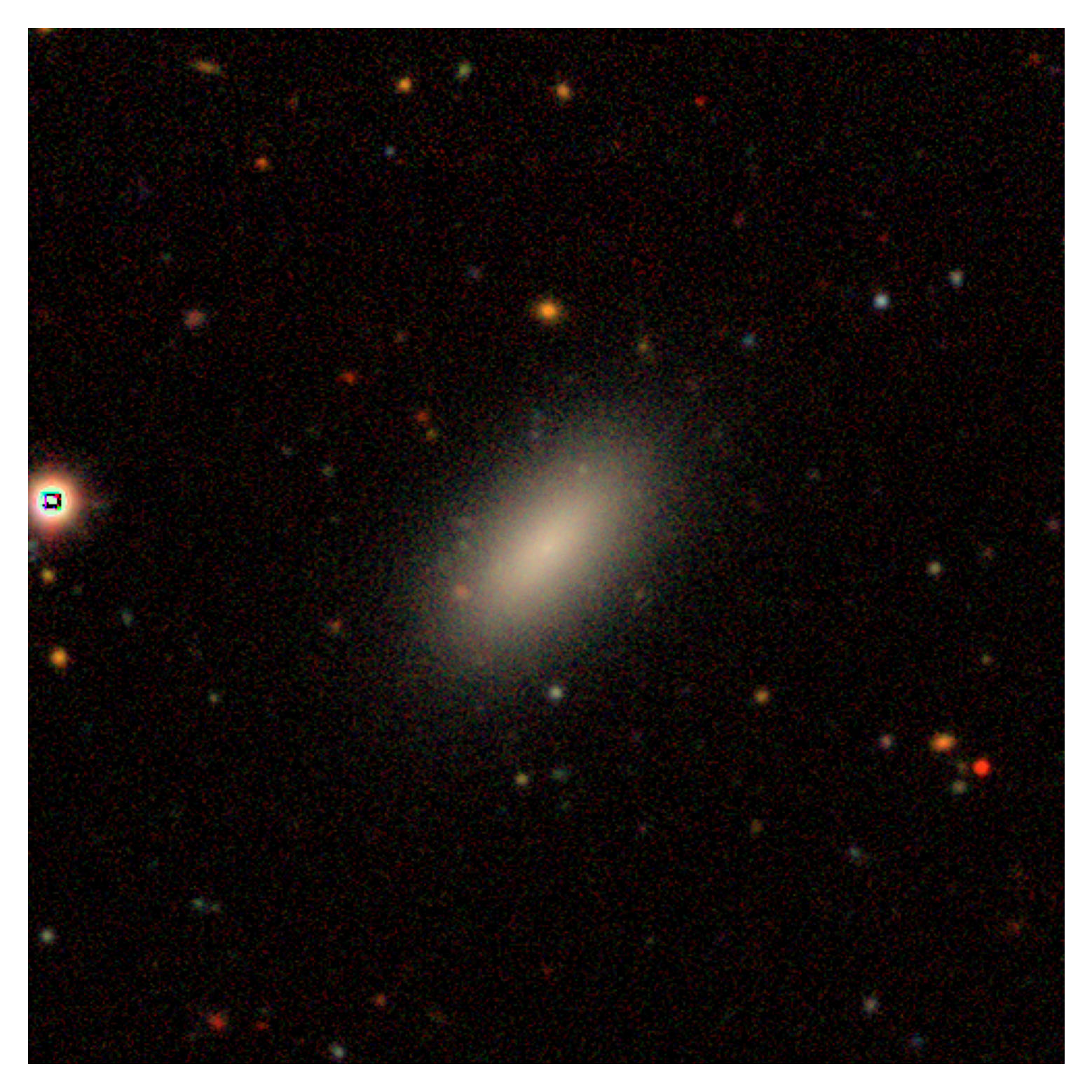}
    \includegraphics[width = 0.3\textwidth]{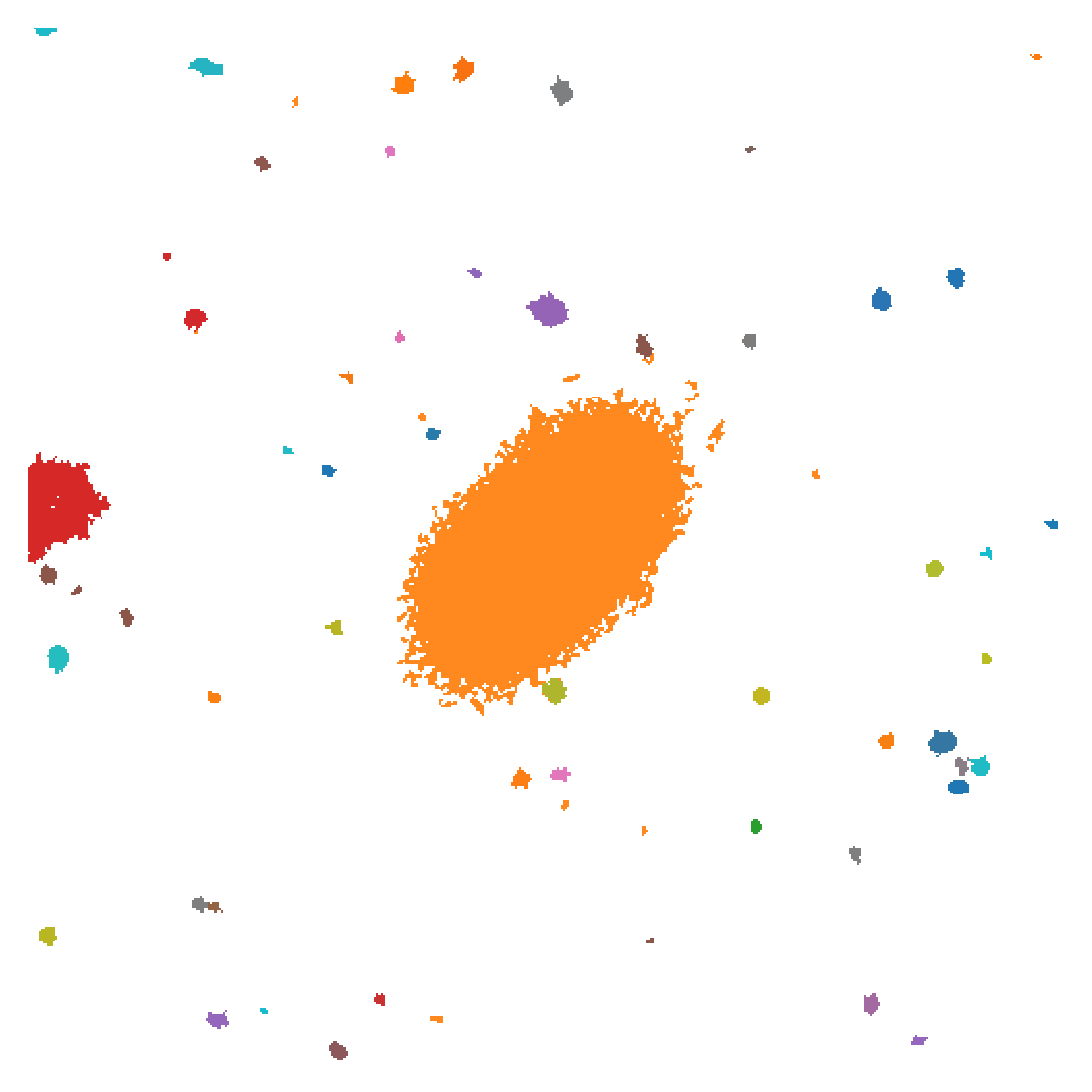}
    \includegraphics[width = 0.3\textwidth]{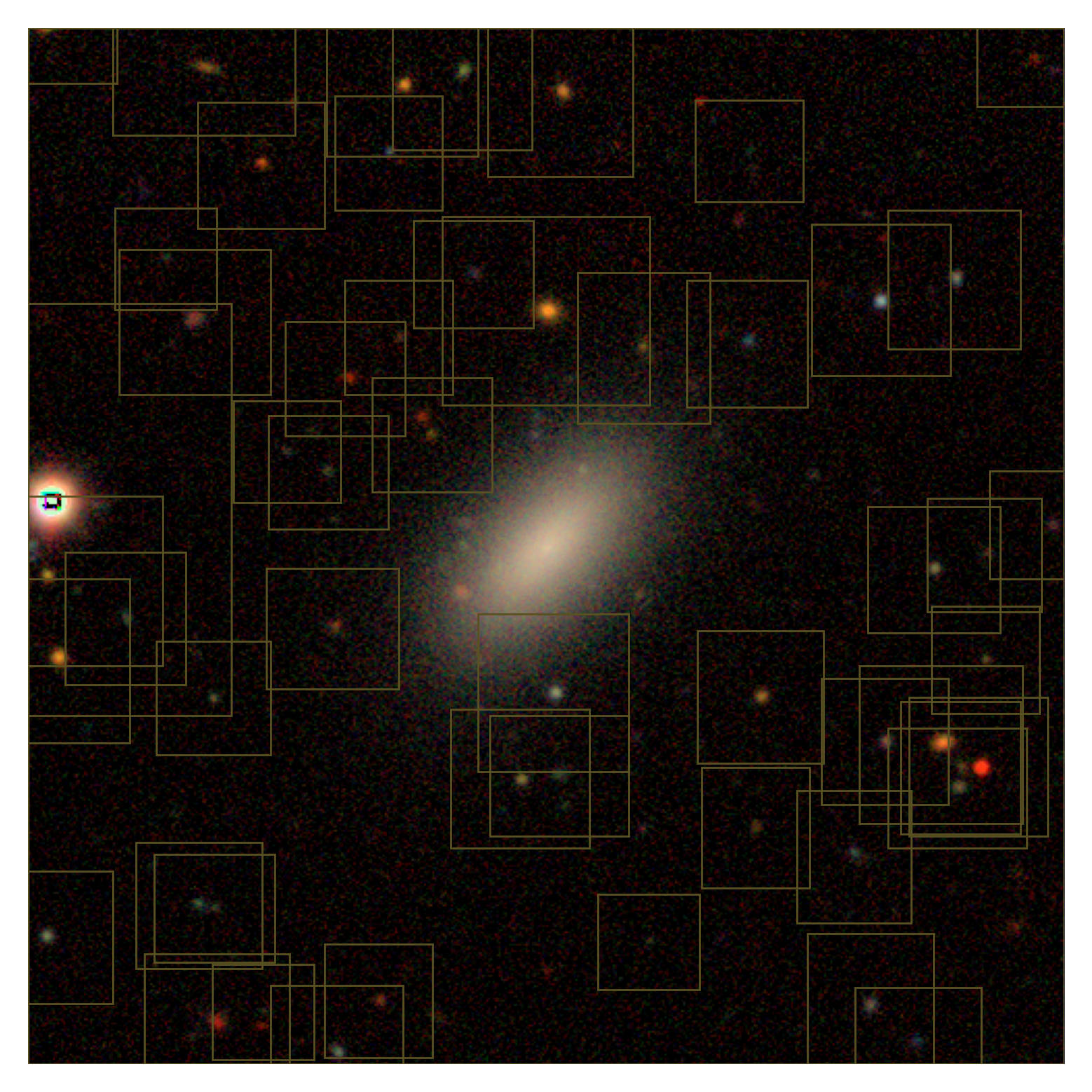}
    \includegraphics[width = \textwidth]{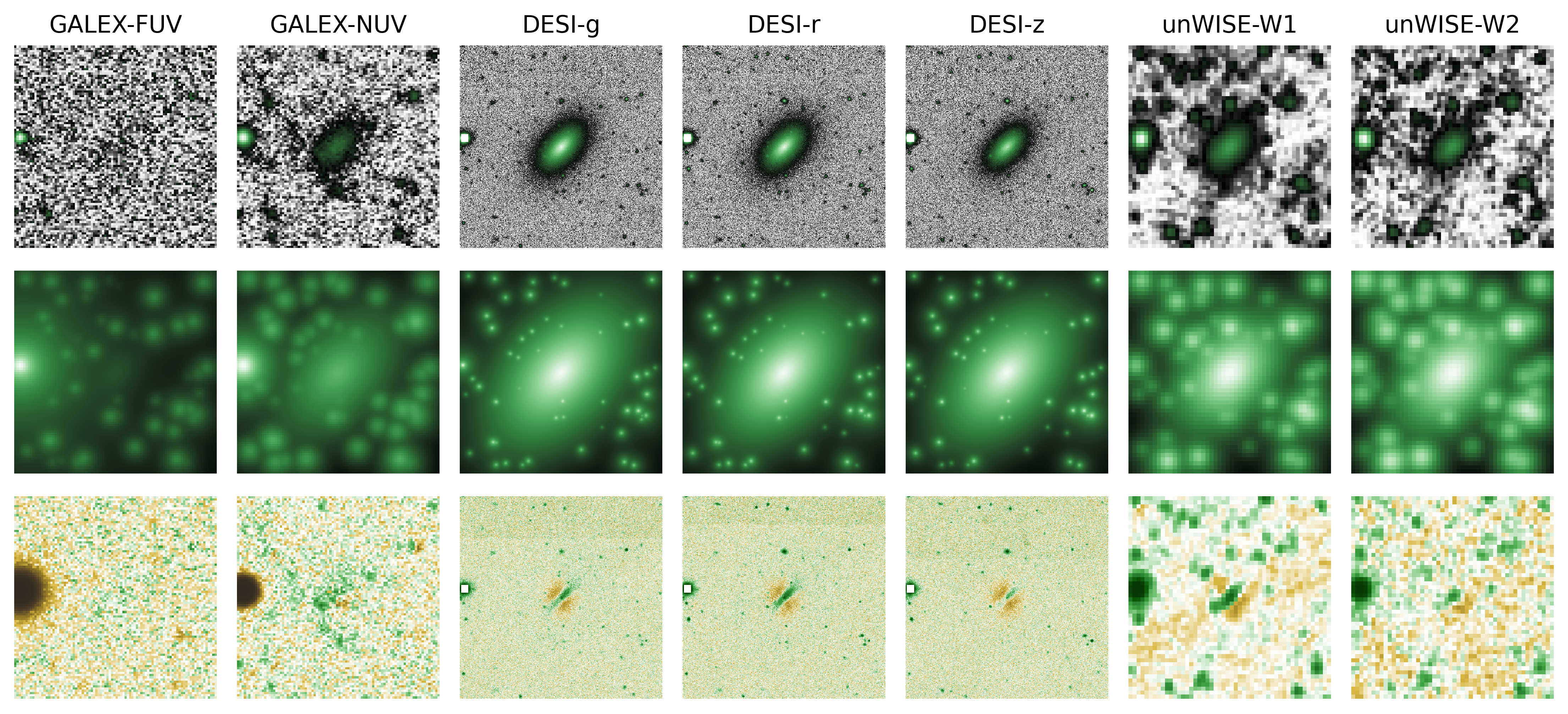}
    \caption{Example of a complex image initialized with a segmentation map and processed with \AP.
    Upper left: Colour image of LEDA~41136 using the g,r,z bands from DESI-LIS~\citep{Dey2019}. 
    Upper middle: Segmentation map generated with {\scriptsize Source Extractor} locates all objects visible in the image. 
    Upper right: Windows in which each model is evaluated.
    Bottom: Initial model automatically generated by \AP for each object in each band (the models are imperfect but a reasonable starting point). 
    The top row is the image in each band, the middle row is the initial model, the bottom row is the initial residuals where green indicates data excess brightness and brown indicates model excess.
    The bottom figure was generated using \AP's automatic plotting routines.
    }
    \label{fig:segmapfit}
\end{figure*}

\Fig{segmapfit} demonstrates this process with LEDA~41136, a dwarf elliptical galaxy with many stars projected onto the same line of sight.
The figure shows the process of creating a segmentation map, here using SourceExtractor~\citep{Bertin1996} with default parameters (except the bright star on the left which was manually fixed due to shredding). 
In total, 54 objects are detected in the image and added to the \AP model seven times each (once for each band) for the various instruments/telescopes.
This segmentation map serves as input to \AP which then constructs windows minimally enclosing each segment.
These windows are then expanded by a factor of two and enlarged with a border of 20 more pixels to ensure all light is included in the window into the LSB regime.
\AP automatically initializes every object in the field with reasonable starting position, ellipticity, position angle (PA), and other parameters.
\AP then fits the parameters and produces the resulting image (see \Sec{examplemultiband}).
Note that we fit an image with a single prominent galaxy here for demonstration purposes, \AP can operate on an arbitrary number of galaxies and stars.

\subsection{Optimization methods}
\label{sec:optimization}

Optimizing model parameters is the core functionality of any astronomical image modelling code. 
However, the specific science goals often dictate the ideal optimization algorithm.
For standard image modelling based on finding the $\chi^2$ minimum (or maximum likelihood) of a given model, \AP includes a custom implementation of the LM algorithm~\citep{Marquardt1963,Transtrum2012,Gavin2022}.
This algorithm smoothly transitions between a gradient descent optimizer and a second order Gauss-Newton's method optimizer; it can handle highly non-linear astronomical image fitting and still obtain high precision parameters. 
The LM algorithm is the method of choice for most astronomical image modelling codes, though several varieties of the algorithm which behave slightly differently also exist~\citep{Gavin2022}.
The \AP implementation of LM is uniquely designed for stability in highly covariant problems with a large number of parameters. 
Unlike other implementations, it can never encounter a ``singular matrix error''. 
Furthermore, the capacity to add arbitrary fuzzy constraints is also available; this is done by treating the constraints as ``pixels'' which can be fit using model parameters.
With a highly flexible constraint system, it is even possible to incorporate other measurements into the fit such as kinematic information that will be modelled simultaneously along with the image.
Given that the LM algorithm is based on an approximation of the Hessian matrix, 
its inverse is used to set uncertainties for the optimal parameter values.
The full covariance matrix (inverse of the Hessian) can be accessed after optimization for a more complete representation of the local uncertainty at the maximum likelihood point.
The main drawback of the LM method is the memory consumption, which is of order $N_{\rm pixels}*N_{\rm parameters}$; for large images with detailed models, this can be prohibitive.

To address memory issues, \AP uses iterative fitters which break the optimization problem into smaller more tractable sub-problems.
This can benefit highly non-linear problems which elude complete optimization, as well as massive optimizations that cannot fit in memory.
Two iterative methods are currently available.
The first method optimizes one sub-model at a time until convergence.
This is slower than a full LM optimization, though its memory requirement is simply that of the largest single sub-model instead of a full LM optimization which scales with the size of the full model.
The second method breaks the list of parameters into ``chunks'' and optimizes only the subset of parameters.
The user can control the structure of these chunks, or allow \AP to randomly select parameters.
The code loops until convergence.
The second method is needed in cases where many models are constrained to share parameters (see \Sec{examplepsfextract}), while the first method is fast and simple to apply if all models are independent.

In some cases, a complete understanding of the errors and any multi-modal properties of the image model is as important as the model parameters themselves.
The typical method of choice for detailed uncertainty modelling is a Metropolis Hastings Markov-Chain-Monte-Carlo (MCMC) and its associated techniques (e.g., annealing).
However, these methods can be slow for high dimensional problems due to the low likelihood of a given proposal step being accepted. 
This causes the convergence time and auto-correlation length to significantly increase.
To overcome this issue, \AP implements the Hamiltonian Monte-Carlo (HMC\footnote{Sometimes referred to as the Hybrid Monte-Carlo.}) method~\citep{Duane1987,Betancourt2011}.
Using information about the gradient of the $\chi^2$, proposing large jumps in parameter space becomes possible with high likelihood, while maintaining detailed balance.
This significantly decreases auto-correlation between samples (no thinning needed), making the chain much more efficient at representing the posterior than the standard Metropolis-Hastings MCMC.
Under typical circumstances, LM can be used to find a high likelihood initialization point for HMC; it is then possible to keep the entire chain (no burn-in required) as the posterior.
A sophisticated variant of HMC called the No-UTurn Sampler (NUTS) can also run on \AP.
Both HMC and NUTS are implemented by the external package {\scriptsize PYRO}~\citep{Bingham2019} which is a powerful tool for differentiable probabilistic programming.
Since \AP models are differentiable, one may easily access the full {\scriptsize PYRO} functionality, though simple wrappers for HMC and NUTS are built-in to \AP.

One can also use a standard gradient descent optimization, of which several options are available through \PT.
The standard method for such cases is the Adam gradient descent~\citep{Kingma2014} which includes momentum in the first and second moments of the gradient.
This and other gradient descent methods can typically converge quickly to approximate solutions. 
However, they do not possess the quadratic convergence property of second order methods like LM and will therefore require far more time to achieve the level of precision typically required for astronomical image analysis.

In \AP, all methods for sampling an image, determining the $\chi^2$ and its gradient, and updating model parameters are exposed to the user.
It is thus possible to implement custom optimizers as well.
However, the broad range of problems that can in principle be tackled by the methods listed above is already quite extensive.

\section{Example \AP Analysis}
\label{sec:examples}

\subsection{Spline Modelling}
\label{sec:examplespline}

The detailed spline models available in \AP allow for a radially varying spline representation of the brightness, PA, and/or axis ratio; along with their corresponding uncertainty.

\begin{figure*}
    \centering
    \includegraphics[width=0.78\textwidth]{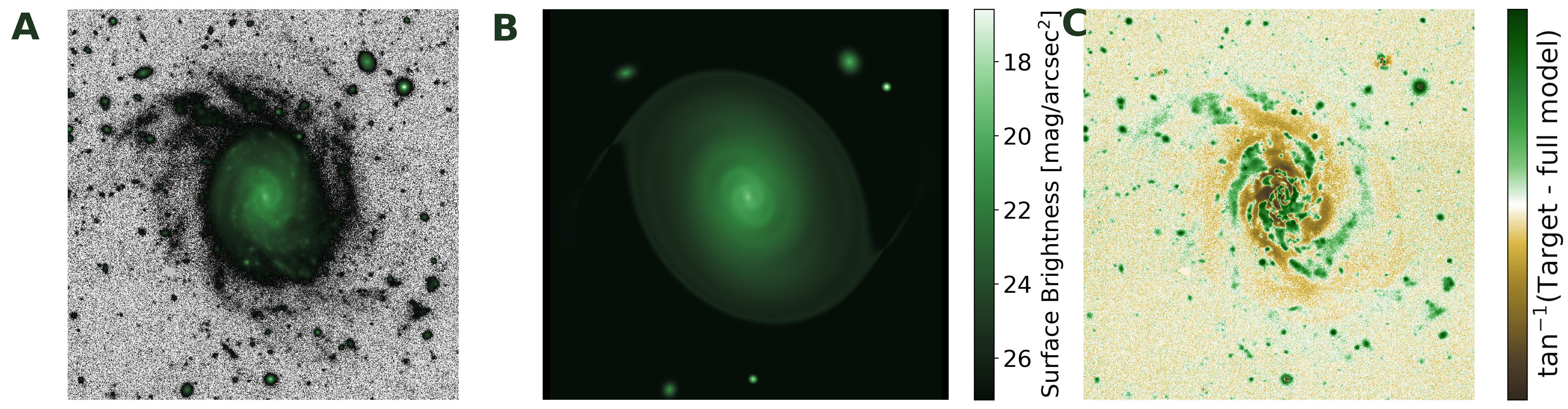}
    \includegraphics[width=0.21\textwidth]{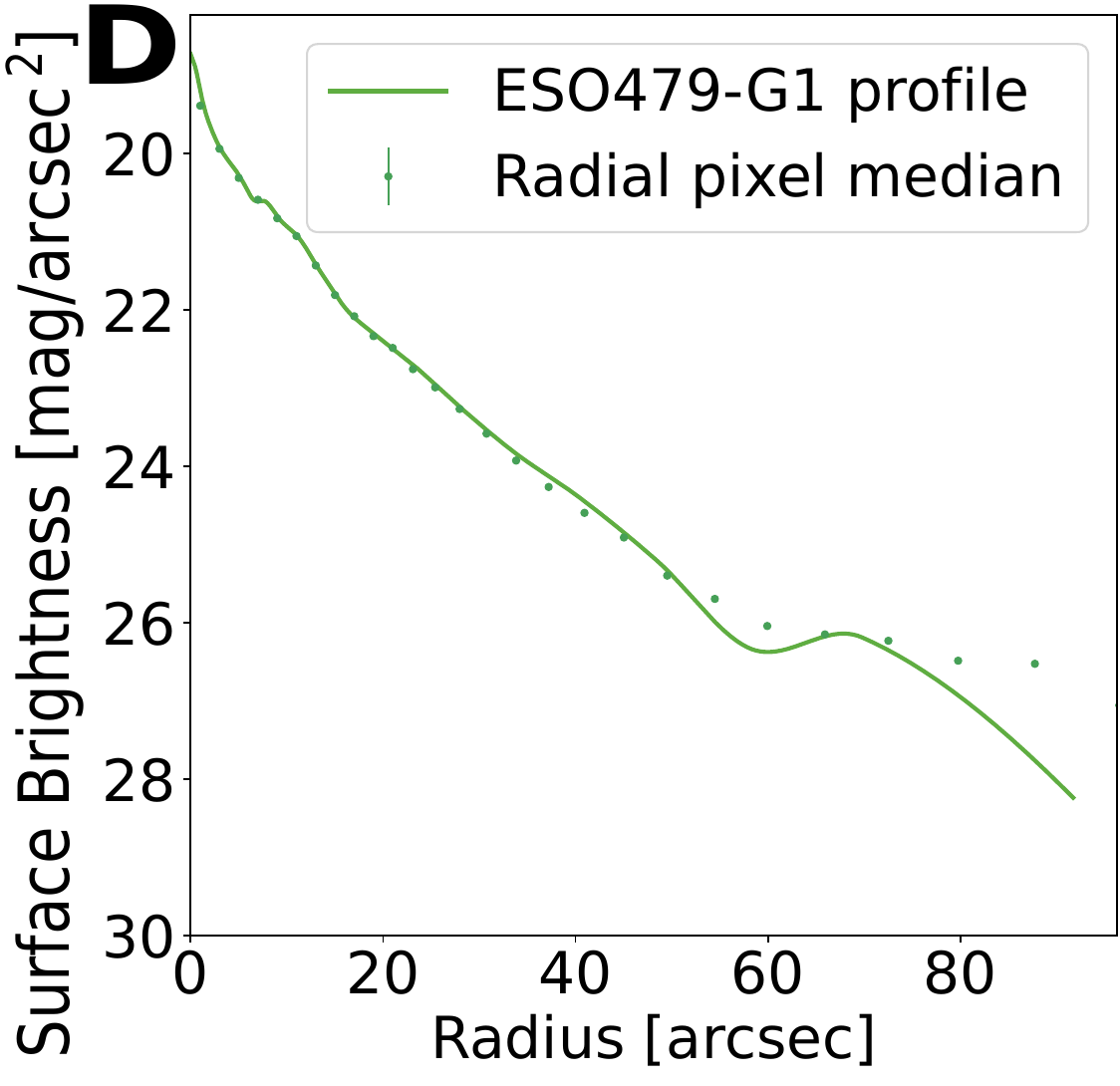}
    \caption{Spline warp galaxy model fit for ESO~479-G1. 
    The panels show A: the r-band DESI-Legacy Survey image of ESO~479-G1; 
    B: the fitted spline model including a few of the brighter nearby objects; 
    C: the data - model residuals, where the colour bar shows data excess (green) and model excess  (brown); 
    and D: the extracted surface brightness profile along with data taken as the median of pixels near each radius. 
    The bar and spiral arms are recovered giving a more representative model of the image. 
    The profile extends to faint surface brightness levels due to the accurate model of the image.
    This profile represents results from a full 2D treatment, compared to the original A{\scriptsize UTO}P{\scriptsize ROF} ellipse fitting tool which worked in 1D.}
    \label{fig:spline}
\end{figure*}

\Fig{spline} demonstrates an example spline warp galaxy model fit to ESO~479-G1, a nearby spiral galaxy.
Clearly visible in the model are a bar and inner spiral arms, demonstrating the expressive power of this numerical representation.
The SB profile is quite deep, much like that shown in Figure 6 of \citet{Stone2021b} using the previous version of \AP.
However, unlike the previous version, \AP can model the other objects in the image simultaneously to achieve greater depth on the image for the main galaxy.
Within this framework, one can also fit detailed spline models for overlapping galaxies.

It took approximately 5 minutes for the model to converge for \Fig{spline} on an \emph{NVIDIA A100} GPU.
Since this is a spline warp model, the light profile, PA and axis ratio all are represented by splines meaning the number of parameters grows quickly.
In this case there are 115 parameters in total which are optimized on 490,000 pixels (700 by 700), including 25 parameters for each spline, one for the sky, and the parameters for the interloping S{\'e}rsic galaxies.

\subsection{GPU Acceleration}
\label{sec:examplegpu}

As noted in \Sec{pytorch}, \AP can use GPU hardware to significantly accelerate image fitting.
At a high level, rather than performing mathematical operations on a per pixel basis, a GPU can execute an array of mathematical operations simultaneously, operating on many hundreds of pixels at a time.
The CPU and GPU operate asynchronously, meaning that while the GPU completes a given set of operations, the CPU can perform basic tasks and determine the next set of operations.
For large images, the GPU is used at capacity for each operation and \AP can fully utilize both the CPU and GPU.
For small images there is little benefit in using a GPU (it may even be slower) as the bottleneck of communicating between the two devices can hamper the overall process.
A near future upgrade to \AP will include the ability to batch small models together so that the time savings can be accessed even for smaller images.
For intermediate-size images, say $\sim 500$ pixels across, operations will be far faster than achieved with a CPU alone but the full potential of the GPU is still not realized.
For larger images, the communication overhead between the CPU and GPU in use becomes negligible and the largest speed improvements are realized.
The exact number of pixels where these transitions occur varies with hardware and the particular problem at hand.
In general, GPUs should provide significant gains whenever a problem is ``big'' in its computational requirements, for smaller fitting tasks, which are quite common, \AP can run on CPU and will still be competitively fast with other fitting codes.

To demonstrate the actual runtime reduction with GPU acceleration, we perform a basic comparison by fitting overlapping S{\'e}rsic models.
For each test, a number $N_{\rm sersic}$ models were generated with random parameters using {\scriptsize GALFIT} (see \Tab{gputestparams}).
The popular {\scriptsize GALFIT} was selected for this comparison since it is written entirely in \emph{C++}, {\scriptsize GALFIT} operates at high effective efficiency. 
It also benefits from over two decades of testing and validation.
To smooth out differences in sub-pixel integration schemes, we generate the test images at 5 times resolution, then pool the pixels back to the 1000x1000 image size for testing.
White and Poisson-like noise were added to the image using a Normal distribution with $\sigma = \sqrt{1 + f}/10$, where $f$ is the flux in a given pixel, to approximate a basic noise model.
The S{\'e}rsic parameters were then randomly perturbed from their true values by $\sim$10\% (see \Tab{gputestparams}) and given as the initialization to {\scriptsize GALFIT} and \AP (both get the exact same initialization).
The runtime for each algorithm and its accuracy at recovering the true parameter values were recorded.

\begin{table}
    \centering
    \begin{tabular}{c c c c}
        Parameter & Unit & Distribution & Initialization \\\hline
        $X$ & pixels & $\mathcal{U}(50,950)$ & $X + \mathcal{N}(0,2)$ \\
        $Y$ & pixels & $\mathcal{U}(50,950)$ & $Y + \mathcal{N}(0,2)$ \\
        $q$ & b/a & $\mathcal{U}(0.3,0.9)$ & $q \mathcal{N}(1,0.1)$ \\
        $PA$ & deg & $\mathcal{U}(0,180)$ & $PA + \mathcal{N}(0,5)$ \\
        $n$ & - & $\mathcal{U}(0.5,4)$ & $n + \mathcal{N}(0,1)$ \\
        $R_{e}$ & pixels & $10^{\mathcal{U}(1.5,2.5)}$ & $R_{e} \mathcal{N}(1,0.1)$ \\
        $I_{e}$ & flux & $10^{\mathcal{N}(0,0.5)}$ & $I_{e} \mathcal{N}(1,0.1)$ \\
    \end{tabular}
    \caption{Setup parameters for drawing random S{\'e}rsic models to compare runtime performance between CPU and GPU. 
    The ``Parameter'' column indicates the S{\'e}rsic model parameter. 
    The ``Units'' column indicates the units of the parameters when being drawn (for simplicity the image has a pixelscale of $1$). 
    The ``Distribution'' column shows the random distribution from which the true parameter values are drawn ($\mathcal{U}$ and $\mathcal{N}$ for uniform and normal distributions respectively). 
    The ``Initialization'' shows the random distributions which perturb the true values before each algorithm (\AP and {\scriptsize GALFIT}) begins optimization. 
    If appropriate ($q$ and $n$), the initialization values are clipped back to their original range.}
    \label{tab:gputestparams}
\end{table}

\begin{figure}
    \centering
    \includegraphics[width = 0.75\columnwidth]{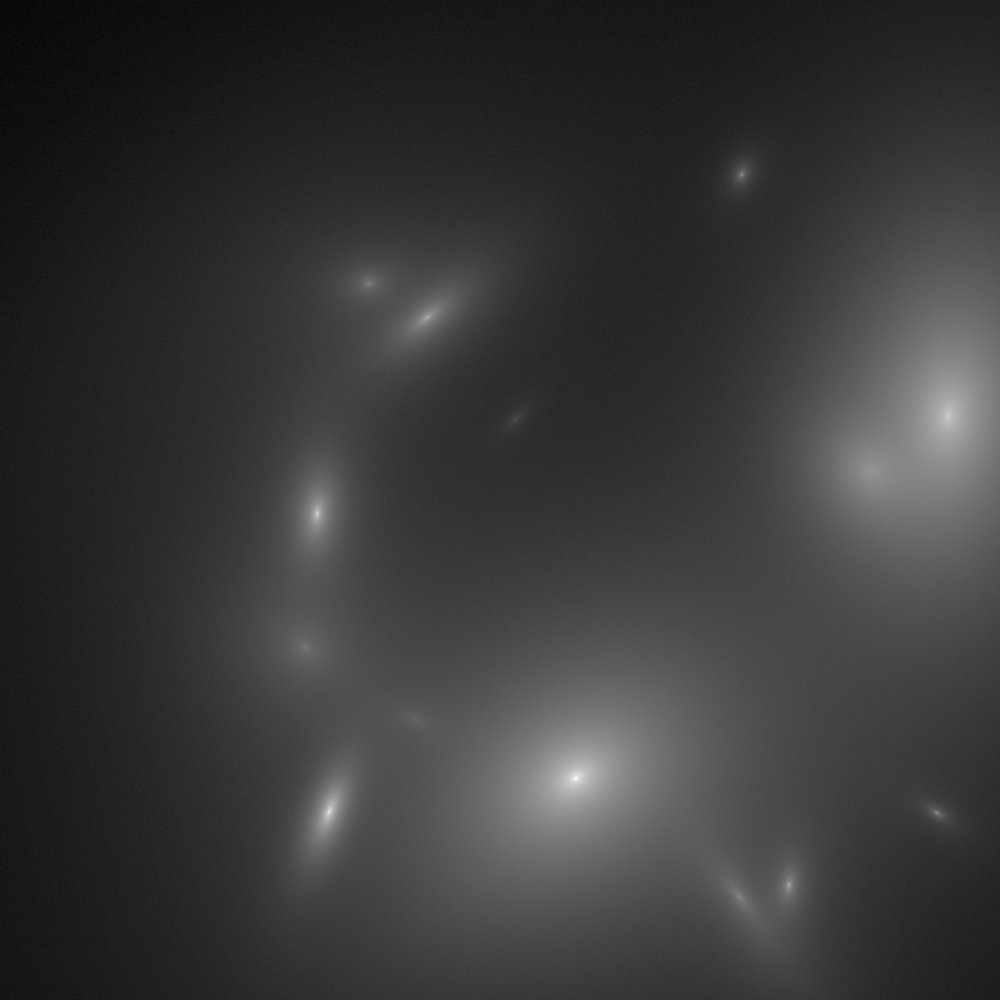}
    \includegraphics[width = 0.95\columnwidth]{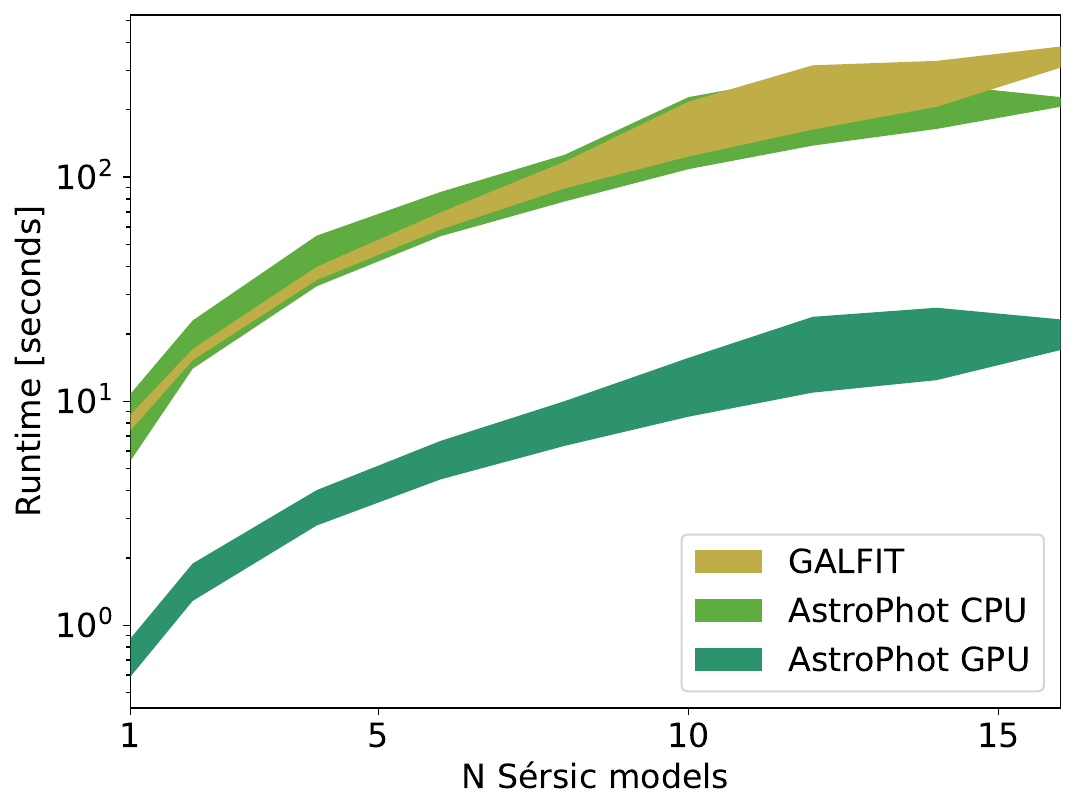}
    \includegraphics[width = 0.95\columnwidth]{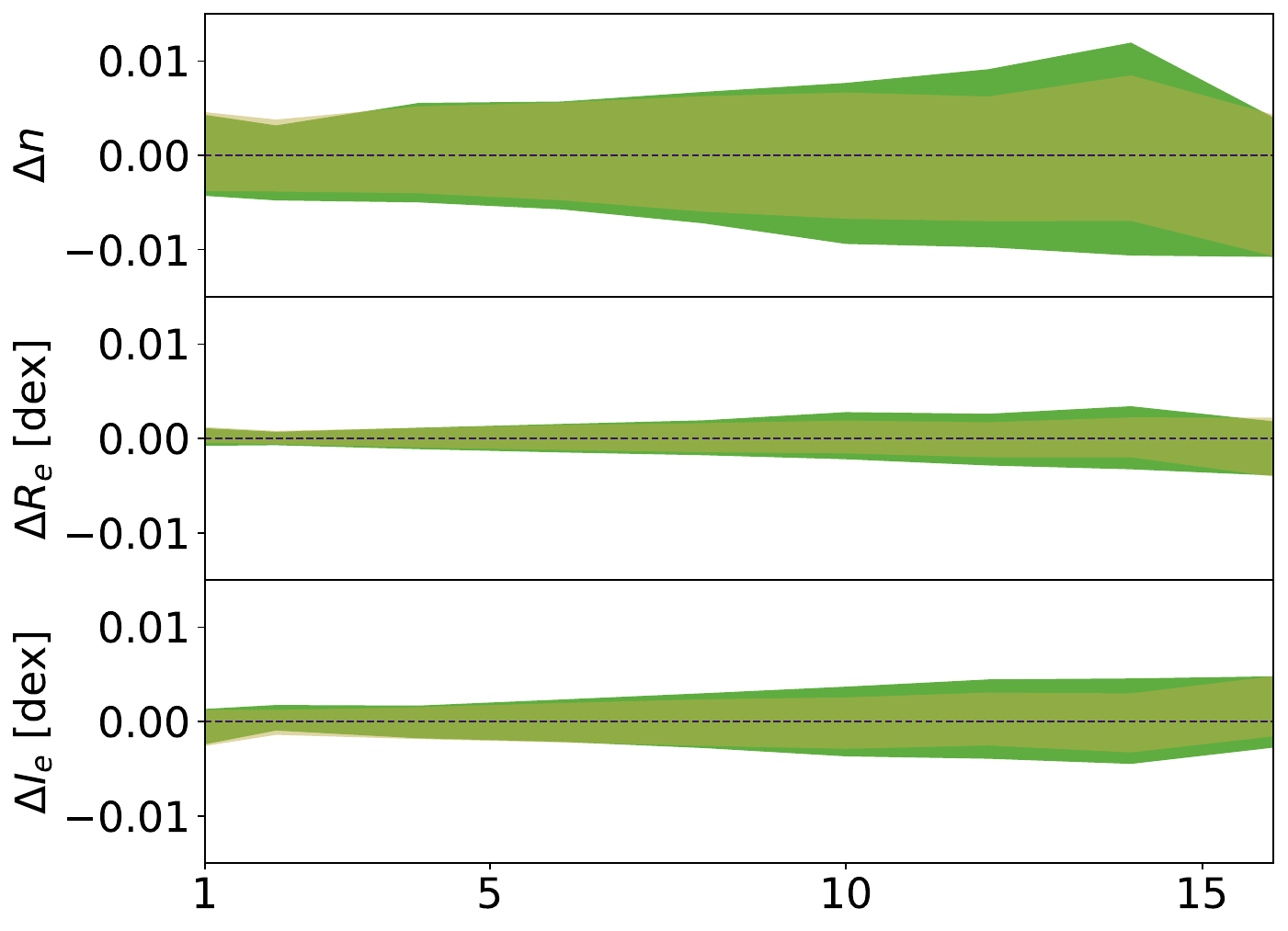}
    \caption{Runtime comparison between \AP and {\scriptsize GALFIT}. 
    Top: example random draw of S{\'e}rsic parameters with noise for 16 galaxies. 
    Middle: relative runtime for \AP and {\scriptsize GALFIT} shown as $16-84$ percentile ranges.
    On CPU, {\scriptsize GALFIT} and \AP have similar runtimes to convergence.
    The \AP-GPU model runs on an A100 GPU and achieves a factor of $\sim$10 speedup over single threaded methods.
    Bottom: relative performance at recovering the core S{\'e}rsic parameters $n$, $R_{e}$, and $I_{e}$ shown as $16-84$ percentile ranges.
    Labels on the y-axis with $\Delta x$ indicate true value minus the model value where $x \in \{n,R_e,I_e\}$.
    \AP shows slightly worse parameter recovery for large models, though the errors are all small.
    }
    \label{fig:runtimegpu}
\end{figure}

\Fig{runtimegpu} shows a comparison between {\scriptsize GALFIT} and \AP.
The relative performance depends on the number of models involved in the fit, as expected since \AP's efficiency improves for larger fits; these require less communication between CPU and GPU and also less communication between the Python script and compiled {\scriptsize PYTORCH} backend. 
\AP on a CPU matches the performance of {\scriptsize GALFIT} with a single CPU, though \PT can do automatic multi-threading on a CPU. 
That is, \AP can be accelerated considerably when more than a single CPU is available.
Most striking is the performance improvement with the GPU where \AP achieves up to a factor of 15 speedup for large models as the GPU's efficiency improves with larger computations.
We caution that direct comparisons between CPU and GPU are not possible; 
the GPU is an \emph{NVIDIA A100} while the CPU is an \emph{AMD Milan 7413}, thus the tests are run on different hardwares.
Similar performance improvements were found on lower tier \emph{NVIDIA V100} GPUs, suggesting that \AP has room for further performance improvements by making better use of the GPU hardware.
Notably, the \AP parameter recovery fared slightly worse on large models than {\scriptsize GALFIT}, however errors were always small and may simply reflect differences in the two codes sub-pixel-integration schemes\footnote{See \Sec{pixelint} and \citet{Robotham2017} for a discussion on sub-pixel-integration.}.
Ultimately, for large pipelines (survey scale analysis) requiring the processing of millions of images, the ``fits per dollar'' calculation is the main driver and this already favors GPU processing~\citep{Sun2019}.
This gap will only increase with time as GPU processing capabilities are increasing exponentially faster than CPUs.
The relative performance with GPU will improve as \AP is further optimized to fully utilize the GPU hardware.
Further relative increases will be available as the \PT community continues to optimize the numerical interface.

\subsection{Multi-band Data Analysis}
\label{sec:examplemultiband}

The next generation of telescopes and large surveys will provide unparalleled multi-band imaging. 
Taking full advantage of these data requires simultaneous fitting of all structures in a given image in a robust and flexible way~\citep{Robotham2022}.
The benefits in signal-to-noise ratio (SNR) boost cannot be ignored as even four bands with similar noise properties provide a factor two improvement in SNR and many surveys will offer even more band-passes per image.
Furthermore, considerable gains can be made by combining the high SNR data and high spatial resolution data (for example) from the multi-facility (e.g., ground- and space-based) imaging of a given object. 
Previous standard solutions such as ``forced PSF-matched photometry'' effectively discard a wealth of available information due to the limitations of single-image fitting.

\begin{figure*}
    \centering
    \includegraphics[width = \textwidth]{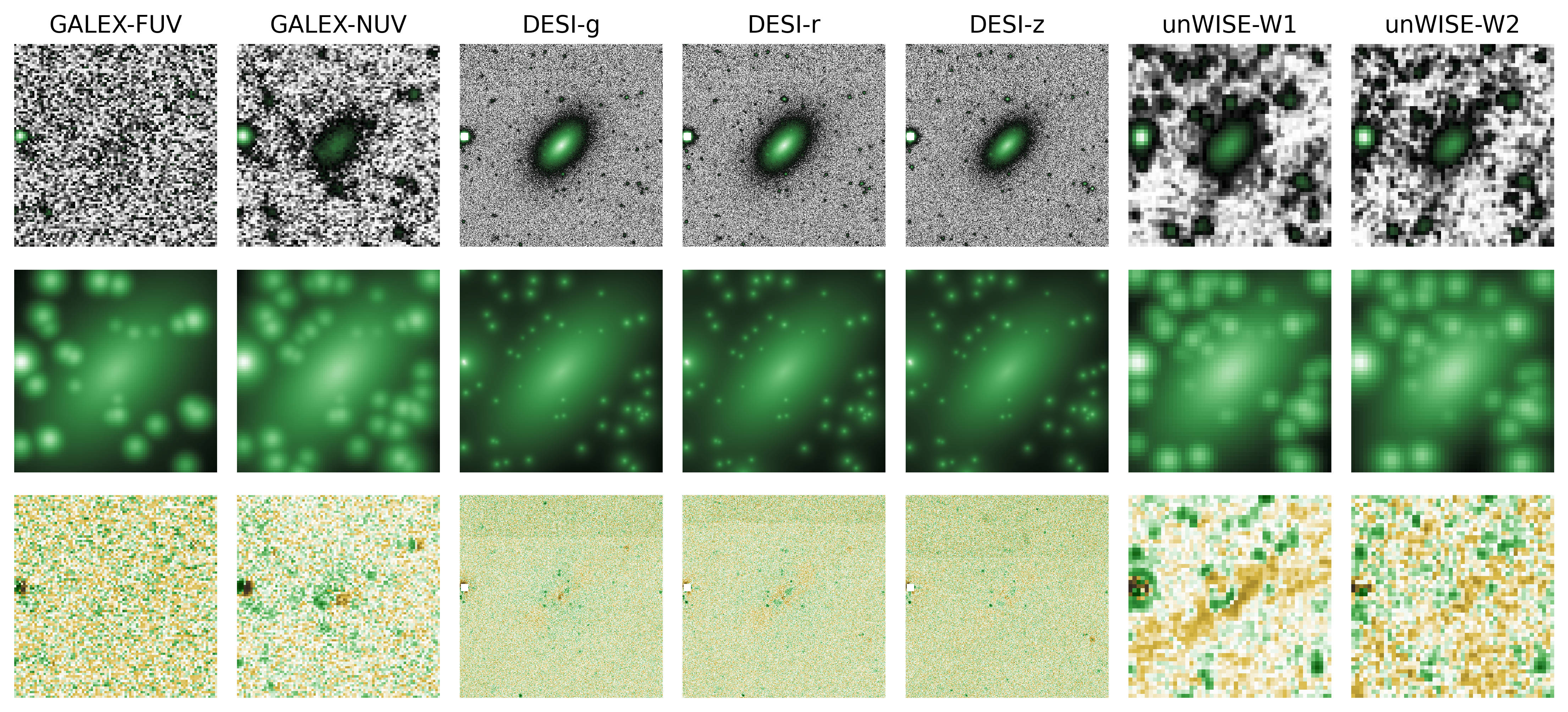}
    \caption{Example of a multi-band data fit using \AP. 
    Here, seven bands are available for LEDA~41136 spanning from the ultraviolet to the mid-infrared each with different PSF, pixel size, and noise characteristics. 
    Top row: image of LEDA~41136 in each of the seven bands. 
    Middle row: fitted model following procedures as in \Sec{intialization}. 
    Bottom row: residual image after subtracting the fitted models (as in \Fig{segmapfit}, brown means the model is brighter than the data), colours are stretched with a \emph{tan$^{-1}$} transform to enhance small values. 
    \AP can synthesize these multi-band data into a unified model which accurately represents the object in all bands simultaneously. 
    All subplots are generated automatically by \AP.
    Note that models appear far more extended in the model images than the observed images due to the lack of noise allowing for a greater dynamic range, though in principle the same information is available in the observed image and the noise simply makes it hard to see.
    }
    \label{fig:multibandfit}
\end{figure*}

\Fig{multibandfit} shows an example multi-band data fit for LEDA~41136 in seven bands: GALEX-FUV, GALEX-NUV, DESI-g, DESI-r, DESI-z, unWISE-W1, and unWISE-W2~\citep[][respectively]{Morrissey2007,Dey2019,Lang2014}.
Each band has a unique PSF model which is optimized concurrently with all other objects in the field (see \Sec{examplepsfextract}).
Similarly, each band has a unique noise model, where the variance is taken to be the flux $16-84$ percentile range (sky noise) plus the sky subtracted flux value for each pixel (Poisson shot noise). 
A S{\'e}rsic model is used for the central object in the image and the $x, y, PA, q, R_{e}, $ and $n$ are constrained to match between bands; however, the brightness ($I_{e}$) is allowed to vary.
All other objects are modelled as Moffat stars with position $x$ and $y$ constrained between bands, and the PSF model ($n$, $R_d$) constrained within a photometric band.
This tightly constrains the structure of all objects in the images; in the FUV band where little structure is seen, these constraints are critical for achieving robust results.
No favoured band is needed in this paradigm, every image contributes to the likelihood in proportion to the amount of signal it contains~\footnote{Forced photometry remains an available option.}.

This is a complex fit involving 771,608 pixels, 54 objects, meaning 378 models, optimized with 504 parameters.
In total it takes 13~min for a single \emph{AMD Rome 7532 2.40 GHz} CPU to converge, or 8~min on an \emph{NVIDIA A100} GPU; this is including the time for \AP's automatic initialization of parameters which take approximately 1~min.
The GPU performs only marginally better than the CPU in this scenario because the setup includes many small objects and GPUs are better optimized for large calculations.
An upcoming update to \AP will include the ability to batch models together, giving access to the full GPU acceleration even for small models.

In multi-band fitting, the accurate measure of the variance for each pixel and the relative on-sky alignment of each image are more critical to make the combined $\chi^2$ meaningful.
The g, r, and z bands still dominate the fit in this instance as they have the highest spatial resolution and SNR.
Information flows from all bands, as can be seen in the bright star near the left edge of the image which is masked in the g-, r-, and z-bands yet is recovered in the fitted model.
For that star the position is constrained by other bands, the PSF model ($n$ and $R_d$) is constrained by other stars in the image, and the brightness is optimized based on the tails of the PSF which are not masked.
For every object in the model, a simultaneous SED is effectively being fit across the multi-band data consistently.
This SED is not a separate constraint or another model to fit, it comes directly from the magnitudes of each component which have been fit in a coherent way such that a full matrix of uncertainty and relative covariance can be accessed for all parameters (see \App{sed}).

Also critical for multi-band fitting is the relative coordinates of the images.
Images must be aligned on a common coordinate grid, as such it is possible to provide the world coordinate position of each image so that they are aligned properly.
Since images from different instruments/telescopes typically have different pixel scales, \AP uses physical coordinates instead of pixel coordinates to track the location of each object\footnote{Setting the pixel scale equal to one effectively sets \AP to pixel units if desired.}.
In fact, a full pixel scale matrix can be used to allow for pixels that are rotated (e.g., multi-epoch imaging with different relative orientations) and pixels that are non-square (any parallelogram is possible).
This can be provided by the user, or one may simply provide an {\scriptsize ASTROPY} \emph{WCS} object~\citep{astropy2022}.
This will automatically handle the relative positioning and pixel scales for multiple images.

The r-band image was used to collect a segmentation map for model initialization.
As such, some objects in the W1 and W2 bands are missing as they were not visible in the r-band image; this can be seen in the W1/W2 residual images.
For a more complete model, image segmentation in multiple bands would be needed, followed by registration of these multiple segmentation images into a single catalogue. 
However, this is beyond the scope of the current paper.

\subsection{Optimal Live PSF Extraction}
\label{sec:examplepsfextract}

A number of techniques have been developed for the extraction of PSF models from astronomical images as they are critical for a number of science objectives~\citep{Krist1993, Jarvis2004, Bertin2011, Lu2017, InfanteSainz2020}.
These typically rely on masking interloping sources, extracting normalized cutouts, re-centering on a pixel, and stacking the resulting images or otherwise modelling the extracted stamps.
While this approach is powerful and can accurately recover PSF models down to very low surface brightness limits, it is labour intensive and operationally challenging in crowded fields.

Rather than stacking, one may use \AP to forward model a PSF during the optimization of other models.
Using shared shape parameters, a global PSF model can be fit along with all other sources in an image.
Each star has a unique brightness and position, while all other parameters are globally fit such that they can be constrained by all stars.
Similarly, the convolution model for extended objects shares the model parameters with the stars, thus fitting a PSF along with all other models.

\begin{figure*}
    \centering
    \includegraphics[width = \textwidth]{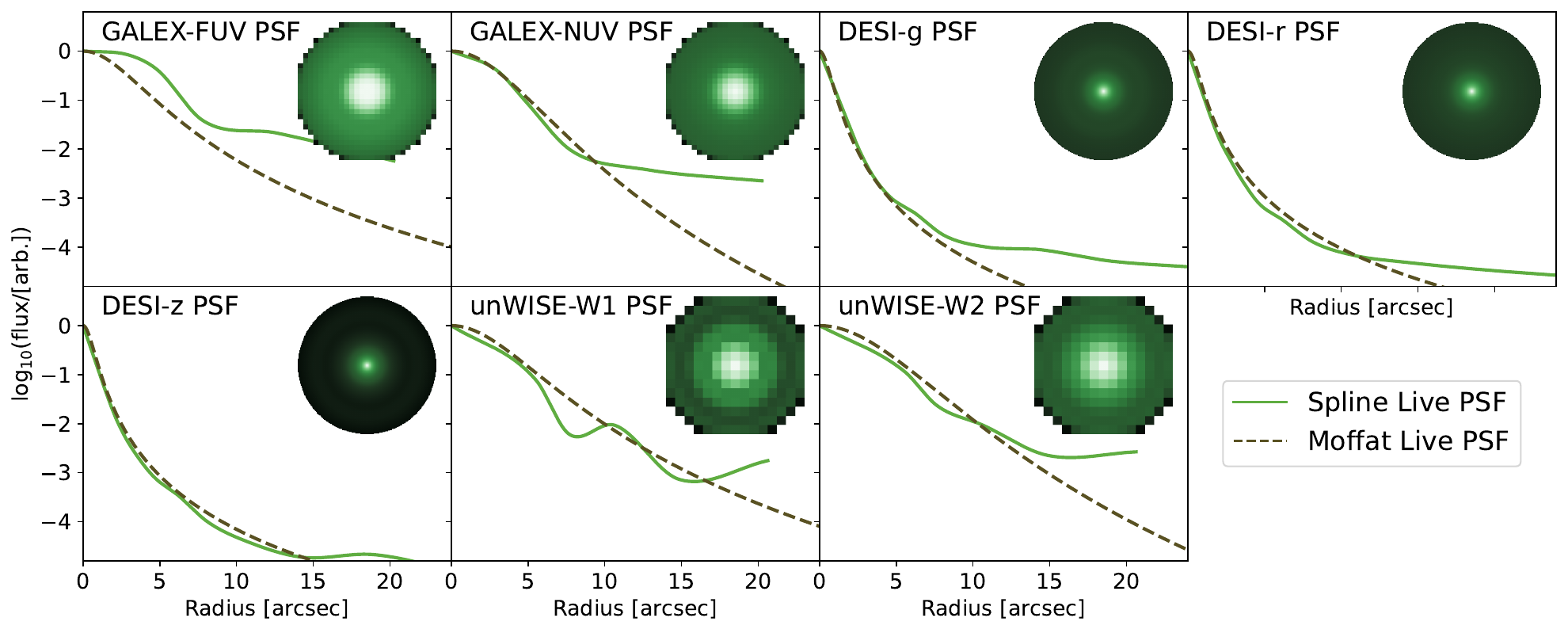}
    \caption{Live PSF models fitted for the multiband data in \Fig{multibandfit}. 
    The y-axis is in arbitrary log flux units for better visualization; the PSFs are normalized internally before convolution. 
    The inset PSF image is for the spline live PSF, the Moffat live PSFs look near identical except smoother. 
    The spline and moffat live PSF models show good agreement to at least two orders of magnitude for all but the FUV-band data which has exceptionally low signal-to-noise.}
    \label{fig:livepsf}
\end{figure*}

\Fig{multibandfit} shows the modelling of many overlapping stars on a single galaxy.
During this fit, the PSF was treated as a model to be optimized along with all the others.
\Fig{livepsf} shows the fitted Moffat PSF model constrained by every object in the image as well as an identical fit except using a spline PSF model\footnote{The spline models are less constrained making the fit considerably more difficult, in total it took 40~min on an A100 GPU to complete the fit.}.
In principle even a single galaxy can be used to model the PSF, and in fact the LEDA4136 galaxy dominates the signal to noise in most bands, likely contributing significantly to constraining the PSF models.

The spline model has a great deal of freedom to fit any radial functional form, it is encouraging to see the Moffat and spline agree to several orders of magnitude in most bands.
Most striking is the z-band data where the agreement is near perfect to over four orders of magnitude. 
The FUV-band PSF shows the least agreement between the spline and Moffat, however the signal-to-noise is low in this band so there is very little constraining power in this image.
In such a scenario it is generally better to use a model like Moffat which effectively imposes a strong prior on the PSF shape.

Interestingly, all of the spline PSF models level off at a higher brightness for large radii than the Moffat PSF.
This could be because at this faint region the spline is unconstrained and so ``freezes'' at an early time in the fit.
More likely, however, is that the extended PSF is in fact brighter than a pure Moffat model.
In these purely circular PSF models we are not accounting for diffraction spikes or other non-axisymmetric features.
The spline may pick up on such faint features, while the Moffat with its pre-described shape is dominated by the bright part of the PSF.
Both unWISE PSFs have noticeable waves in the profile, a likely culprit is that several objects in these bands are not being modelled, causing the PSFs to be slightly perturbed during the optimization.
These extra sources were not identified in the segmentation map on the r-band from \Fig{segmapfit}.
It is common for infrared images to have more point sources that don't show up in optical bands.

\subsection{Faster Bayesian Photometry} 

\label{sec:examplehmc}
While finding optimal parameter values is always a central goal, quantifying the uncertainty of a given fit is equally critical.
A common strategy for exploring the uncertainty of any fit is the Markov-Chain Monte-Carlo (MCMC).
This ubiquitous technique is robust and effective at exploring parameter space, however at the expense of speed.
As such, various techniques have been developed to accelerate the sampling process.
Some of the most sophisticated ones include score (gradient) based sampling methods such as the Hamiltonian Monte-Carlo~\citep{Betancourt2017} and its variant, the No U-Turn Sampling method~\citep[NUTS][]{Hoffman2011}.

NUTS can produce samples with very low auto-correlation in a fraction of the time needed for a vanilla Metropolis-Hastings MCMC. 
This is done by using the gradient of the likelihood to inform more meaningful steps in the parameter space.
This technique treats the probability distribution as a potential in which a particle can move using Hamiltonian dynamics.
By assigning a random initial momentum to the particle and evolving it for several iterations, one can significantly reduce the correlation length between samples whilst maintaining the detailed balance needed for a Markov-Chain.
For comparison, a traditional Metropolis Hastings MCMC typically scales as $\mathcal{O}(D)$ where $D$ is the number of dimensions~\citep{Roberts2014}, whereas the HMC scales as $\mathcal{O}(D^{1/4})$~\citep{Beskos2013} though the practical limits depend on the nature of the problem.

\begin{figure*}
    \centering
    \includegraphics[width = 0.49\textwidth]{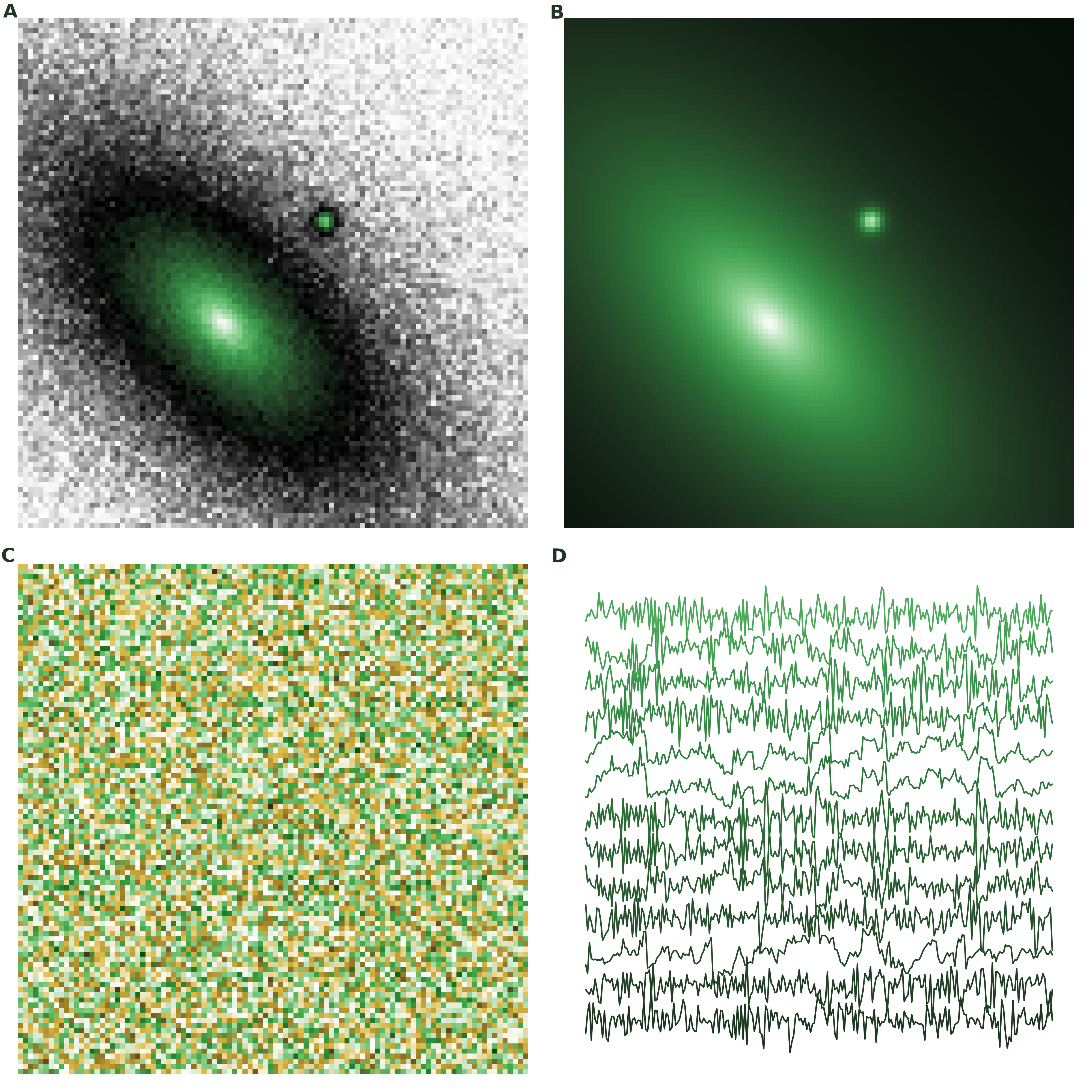}
    \includegraphics[width = 0.49\textwidth]{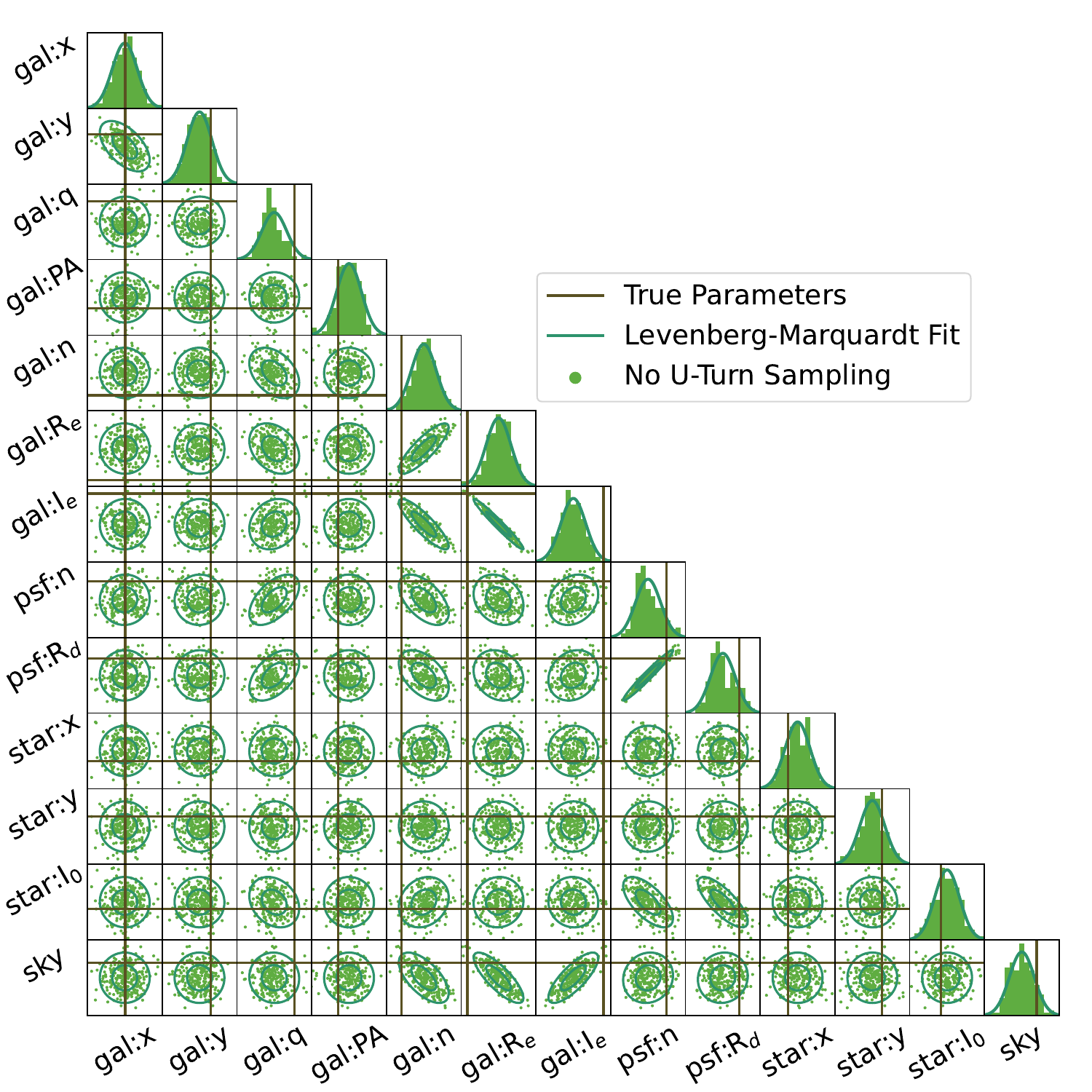}
    \caption{Example of a posterior computed with LM and the gradient based No U-Turn Sampling method for a galaxy, star, sky model with convolved PSF. 
    A: simulated target image with noise to model with both techniques; 
    B: fitted model, which is visually indistinguishable from the true model;
    C: residuals after fitting, which are consistent with random noise;
    D: representation of the sampling chains from NUTS, the chains are visibly well mixed over the 250 samples. The chains from bottom to top have the same order as the parameters in the corner plot.
    Right: Corner plot showing posterior information for all the fitted parameters. 
    Contours give the 68 and 95\% credible regions from the LM method, the scatter points are samples from NUTS. 
    Vertical lines give the true values for each parameter.
    }
    \label{fig:hmcfit}
\end{figure*}

\Fig{hmcfit} shows an example of a model fit on mock data with several interacting features.
It contains a galaxy (S{\'e}rsic), a star (Moffat), and a sky background level, the galaxy is also convolved with a PSF which must also be optimized.
In total there are 13 parameters in this model, seven for the galaxy ($x$, $y$, $q$, $PA$, $n$, $R_e$, $I_e$), two for the PSF ($n$, $R_d$), three for the star ($x$, $y$, $I_0$), and one for the sky\footnote{Note that the star model also uses the two PSF parameters.}.
Noise is added to the image simply as a normal distribution with a standard deviation equal to the square root of the flux divided by 2\footnote{The factor of two is arbitrary and would in reality just be a function of the integration time of the image.}, the final signal to noise is $\sim 8$.
The model is initialized far from the true parameters and then fit with the LM algorithm.
LM takes 3.4 seconds to converge, which is negligible compared to the sampling time.
From the maximum likelihood point found by LM, NUTS begins sampling, using the LM covariance matrix as the mass matrix to improve sampling efficiency.
First a warmup of 20 samples is used to select an optimal step size, then 250 samples are drawn using NUTS.
This sampling took 738.5 seconds to complete, which is quite fast for sampling such a complex parameter space.

\Fig{hmcfit} shows good agreement between the LM covariance matrix and the NUTS samples.
This is in fact quite common for photometric models.
In most scenarios the LM results, which are orders of magnitude faster to obtain, are sufficient to represent the relevant uncertainty in all parameters.
One should only perform sampling in situations where non-linearities are expected to be strong, such as with highly non-linear models and low signal-to-noise.
These situations are quite rare, and instead of the expensive sampling, 
most applications can benefit from the covariance matrix directly from LM, which \AP provides easy access to.

In \Fig{hmcfit}, the true values also appear to be somewhat displaced from the uncertainty limits.
This is most visible for the ``gal:R$_e$'' and ``gal:I$_e$'' parameters where the true values are outside the 95\% credible region.
While the true values may appear far from the fitted values, they are in fact very close.
The fitted R$_e$ is 30.9$^{\prime\prime}$, while the true value is 30$^{\prime\prime}$; similarly the fitted I$_e$ is 0.98 log(flux) while, the true value is 1 log(flux).
It is common for the statistical uncertainties to underestimate the true uncertainty, especially for fits to real data where the fitted models (i.e., S{\'e}rsic) are only approximations to the true shape of a galaxy.
Still, the covariance between parameters is real and informative when examining a fitted model.

As a final remark on Bayesian image analysis, we consider the case where prior information is to be encoded into the fit.
With NUTS, priors can be encoded using the {\scriptsize PYRO} package systems, however \AP currently has no formal system to include prior information during the LM fitting process.
In practice, this is not an issue as photometry is typically rather constraining and the prior contributes little to the posterior.
In this regime where the likelihood is dominating, one may use Bayes theorem more literally:

\begin{equation}
    P(\theta | D) = \frac{P(D | \theta) P(\theta)}{P(D)}, 
\end{equation}

\noindent where $\theta$ are the \AP model parameters and $D$ are the data. 
The likelihood $P(D | \theta)$ is given by \AP's LM as a multivariate Gaussian which is typically straightforward to multiply by the prior.
In the common scenario where the prior is also encoded as a multivariate Gaussian, one can apply Bayes theorem after LM optimization using Bayes theorem for Gaussians~\citep[section 8.1.8 in][]{petersen2008matrix}:

\begin{equation}
\begin{aligned}    
    \Sigma_3 &= (\Sigma_1^{-1} + \Sigma_2^{-1})^{-1} \\
    \theta_3 &= \Sigma_3(\Sigma_1^{-1}\theta_1 + \Sigma_2^{-1}\theta_2),
\end{aligned}
\end{equation}

\noindent where $\Sigma_i$ represents the covariance matrices, $\theta_i$ represents the parameter vectors, and $1,2,3$ index the likelihood, prior, and posterior respectively.
In the domain where the likelihood dominates, this formalism captures all of the desired information.
If the prior is not available as a multivariate Gaussian, the multiplication can easily be performed numerically.

This also suggests an accelerated algorithm for fitting multi-epoch data of the same object.
Fitting images individually with \AP produces a multivariate Gaussian for each image.
One may then take the product of all the output Gaussians which is equivalent to having fitted all of the images simultaneously.
It is computationally and logistically easiest to fit each image individually.
By taking the product of the Gaussians afterwards, one can then propagate all the information properly into the final results. 

\section{Conclusions}
\label{sec:conclusions}

As we enter a new era of massive astronomical imaging programs, processing algorithms must adapt to fully exploit these data.
For large survey programs to meet their science goals will require massive computational resources, calling for a reduction of the ``FLOPS per dollar'' needed to perform astronomical image modeling.
Image processing with \AP represents a clear case where GPU acceleration provides a substantial improvement.
Many other data processing tasks in astronomy can take advantage of GPUs and differentiable programming given the relative ease of \PT programming.
GPUs are designed to execute a large number of simple operations in parallel. 
They are especially well suited to image processing, integration, numerical simulation, and other parallelizable tasks common to astronomy.
The power of automatic differentiation also extends the range of available algorithms for previously prohibitive research.
Taken together, GPU acceleration and automatic differentiation herald a new era of astronomical data processing, just in time for the onslaught of new multi-band astronomical images.

In this paper, we have demonstrated the capabilities of the \AP image analysis package and highlighted its value for the next generation of science goals and astronomical imaging programs.
When powered by the \PT package, originally designed to accelerate ML training, we also showed that \AP performs $\chi^2$ minimization quickly and efficiently. 
While one cannot compare CPUs and GPUs exactly, we found in \Sec{examplegpu} typical accelerations of order-of-magnitude faster with GPUs using \AP compared to state-of-the-art single CPU processing.
We also showed in \Sec{examplemultiband} that \AP can take full advantage of multi-band data collected from multiple telescopes with widely different characteristics to extract all available information.
This demonstration shows the broad capability of \AP to fit galaxies and stars in images of any size under a broad range of observing conditions.
In \Sec{examplepsfextract} we discussed the powerful capability of \AP to treat PSF parameters like other model parameters, allowing ``live'' PSF fitting along with all other models.
Finally, we discussed in \Sec{examplehmc} how automatic differentiation opens new avenues for Bayesian analysis with the HMC algorithm. 
Robust posteriors can now be acquired quickly even for complex models of astronomical objects.

Overall, \AP is well suited for a wide range of astronomical research by providing detailed representations for everything, everywhere, all at once in astronomical images from massive catalogues to singular studies of objects at any shape or distance.
With a robust base and flexible extension, \AP can tackle a myriad of tasks previously considered prohibitive and can fully take advantage of available resources both computationally and scientifically.
Astrometry, precision photometry, PSF modelling, globular clusters, ultra-diffuse galaxies, spiral galaxies, elliptical galaxies, diffuse stellar streams, massive galaxy clusters, intra cluster light, cosmological studies, and everything in-between can now be tackled with \AP.

\section*{Data Availability}
The data underlying this article will be shared on reasonable request to the corresponding author(s).

\section*{Acknowledgements}
We gratefully acknowledge judicious comments by the referee, Aaron Robotham; these included, among others, the excellent suggestion to expand our discussion on ``live PSF fitting''
(\Sec{examplepsfextract}).
Henri Gavin and Chien Peng are also thanked for advice on the implementation of the Levenberg-Marquardt algorithm and helpful comments on the presentation of the paper.
This research has made use of the NASA/IPAC Extragalactic Database (NED), which is funded by the National Aeronautics and Space Administration and operated by the California Institute of Technology.

CS acknowledges the support of a NSERC Postdoctoral Fellowship and a CITA National Fellowship.
SC is grateful to the Natural Sciences and Engineering Research Council of Canada and Queen's University for their extensive and generous support.
This research was enabled in part by support provided by Calcul Qu\'ebec, the Digital Research Alliance of Canada, and a generous donation by Eric and Wendy Schmidt with the recommendation of the Schmidt Futures Foundation. 
Y.H. and L.P. acknowledge support from the National Sciences and Engineering Council of Canada grants RGPIN-2020-05073 and 05102, the Fonds de recherche du Québec grants 2022-NC-301305 and 300397, and the Canada Research Chairs Program.
NA thanks the support of Tamkeen under the New York University Abu Dhabi Research Institute grant CAP$^3$.
We acknowledge the use of the GPT-4 model by OpenAI for providing guidance during the programming of \AP and editing of the paper.

\bibliographystyle{mnras}
\bibliography{AstroPhot}

\begin{thebibliography}{}
\makeatletter
\relax
\def\mn@urlcharsother{\let\do\@makeother \do\$\do\&\do\#\do\^\do\_\do\%\do\~}
\def\mn@doi{\begingroup\mn@urlcharsother \@ifnextchar [ {\mn@doi@}
  {\mn@doi@[]}}
\def\mn@doi@[#1]#2{\def\@tempa{#1}\ifx\@tempa\@empty \href
  {http://dx.doi.org/#2} {doi:#2}\else \href {http://dx.doi.org/#2} {#1}\fi
  \endgroup}
\def\mn@eprint#1#2{\mn@eprint@#1:#2::\@nil}
\def\mn@eprint@arXiv#1{\href {http://arxiv.org/abs/#1} {{\tt arXiv:#1}}}
\def\mn@eprint@dblp#1{\href {http://dblp.uni-trier.de/rec/bibtex/#1.xml}
  {dblp:#1}}
\def\mn@eprint@#1:#2:#3:#4\@nil{\def\@tempa {#1}\def\@tempb {#2}\def\@tempc
  {#3}\ifx \@tempc \@empty \let \@tempc \@tempb \let \@tempb \@tempa \fi \ifx
  \@tempb \@empty \def\@tempb {arXiv}\fi \@ifundefined
  {mn@eprint@\@tempb}{\@tempb:\@tempc}{\expandafter \expandafter \csname
  mn@eprint@\@tempb\endcsname \expandafter{\@tempc}}}

\bibitem[\protect\citeauthoryear{{Abraham} \& {van Dokkum}}{{Abraham} \& {van
  Dokkum}}{2014}]{Abraham2014}
{Abraham} R.~G.,  {van Dokkum} P.~G.,  2014, \mn@doi [\pasp] {10.1086/674875},
  \href {https://ui.adsabs.harvard.edu/abs/2014PASP..126...55A} {126, 55}

\bibitem[\protect\citeauthoryear{{Adam}, {Coogan}, {Malkin}, {Legin},
  {Perreault-Levasseur}, {Hezaveh}  \& {Bengio}}{{Adam}
  et~al.}{2022}]{Adam2022}
{Adam} A.,  {Coogan} A.,  {Malkin} N.,  {Legin} R.,  {Perreault-Levasseur} L.,
  {Hezaveh} Y.,   {Bengio} Y.,  2022, \mn@doi [arXiv e-prints]
  {10.48550/arXiv.2211.03812}, \href
  {https://ui.adsabs.harvard.edu/abs/2022arXiv221103812A} {p. arXiv:2211.03812}

\bibitem[\protect\citeauthoryear{{Adam}, {Perreault-Levasseur}, {Hezaveh}  \&
  {Welling}}{{Adam} et~al.}{2023}]{Adam2023}
{Adam} A.,  {Perreault-Levasseur} L.,  {Hezaveh} Y.,   {Welling} M.,  2023,
  \mn@doi [arXiv e-prints] {10.48550/arXiv.2301.04168}, \href
  {https://ui.adsabs.harvard.edu/abs/2023arXiv230104168A} {p. arXiv:2301.04168}

\bibitem[\protect\citeauthoryear{{Akeson} et~al.,}{{Akeson}
  et~al.}{2019}]{Akeson2019}
{Akeson} R.,  et~al., 2019, \mn@doi [arXiv e-prints]
  {10.48550/arXiv.1902.05569}, \href
  {https://ui.adsabs.harvard.edu/abs/2019arXiv190205569A} {p. arXiv:1902.05569}

\bibitem[\protect\citeauthoryear{{Astropy Collaboration} et~al.,}{{Astropy
  Collaboration} et~al.}{2022}]{astropy2022}
{Astropy Collaboration} et~al., 2022, \mn@doi [\apj]
  {10.3847/1538-4357/ac7c74}, \href
  {https://ui.adsabs.harvard.edu/abs/2022ApJ...935..167A} {935, 167}

\bibitem[\protect\citeauthoryear{{Barden}, {H{\"a}u{\ss}ler}, {Peng},
  {McIntosh}  \& {Guo}}{{Barden} et~al.}{2012}]{Barden2012}
{Barden} M.,  {H{\"a}u{\ss}ler} B.,  {Peng} C.~Y.,  {McIntosh} D.~H.,   {Guo}
  Y.,  2012, \mn@doi [\mnras] {10.1111/j.1365-2966.2012.20619.x}, \href
  {https://ui.adsabs.harvard.edu/abs/2012MNRAS.422..449B} {422, 449}

\bibitem[\protect\citeauthoryear{{Bertin}}{{Bertin}}{2011}]{Bertin2011}
{Bertin} E.,  2011, in {Evans} I.~N.,  {Accomazzi} A.,  {Mink} D.~J.,   {Rots}
  A.~H.,  eds,  Astronomical Society of the Pacific Conference Series Vol. 442,
  Astronomical Data Analysis Software and Systems XX. p.~435

\bibitem[\protect\citeauthoryear{{Bertin} \& {Arnouts}}{{Bertin} \&
  {Arnouts}}{1996}]{Bertin1996}
{Bertin} E.,  {Arnouts} S.,  1996, \mn@doi [\aaps] {10.1051/aas:1996164}, \href
  {https://ui.adsabs.harvard.edu/abs/1996A&AS..117..393B} {117, 393}

\bibitem[\protect\citeauthoryear{{Bertin}, {Mellier}, {Radovich}, {Missonnier},
  {Didelon}  \& {Morin}}{{Bertin} et~al.}{2002}]{Bertin2002}
{Bertin} E.,  {Mellier} Y.,  {Radovich} M.,  {Missonnier} G.,  {Didelon} P.,
  {Morin} B.,  2002, in {Bohlender} D.~A.,  {Durand} D.,   {Handley} T.~H.,
  eds,  Astronomical Society of the Pacific Conference Series Vol. 281,
  Astronomical Data Analysis Software and Systems XI. p.~228

\bibitem[\protect\citeauthoryear{{Bertin}, {Schefer}, {Apostolakos},
  {{\'A}lvarez-Ayll{\'o}n}, {Dubath}  \& {K{\"u}mmel}}{{Bertin}
  et~al.}{2020}]{Bertin2020}
{Bertin} E.,  {Schefer} M.,  {Apostolakos} N.,  {{\'A}lvarez-Ayll{\'o}n} A.,
  {Dubath} P.,   {K{\"u}mmel} M.,  2020, in {Pizzo} R.,  {Deul} E.~R.,  {Mol}
  J.~D.,  {de Plaa} J.,   {Verkouter} H.,  eds,  Astronomical Society of the
  Pacific Conference Series Vol. 527, Astronomical Data Analysis Software and
  Systems XXIX. p.~461

\bibitem[\protect\citeauthoryear{Beskos, Pillai, Roberts, Sanz-Serna  \&
  Stuart}{Beskos et~al.}{2013}]{Beskos2013}
Beskos A.,  Pillai N.,  Roberts G.,  Sanz-Serna J.-M.,   Stuart A.,  2013,
  \mn@doi [Bernoulli] {10.3150/12-BEJ414}, 19, 1501

\bibitem[\protect\citeauthoryear{{Betancourt}}{{Betancourt}}{2017}]{Betancourt2017}
{Betancourt} M.,  2017, \mn@doi [arXiv e-prints] {10.48550/arXiv.1701.02434},
  \href {https://ui.adsabs.harvard.edu/abs/2017arXiv170102434B} {p.
  arXiv:1701.02434}

\bibitem[\protect\citeauthoryear{{Betancourt} \& {Stein}}{{Betancourt} \&
  {Stein}}{2011}]{Betancourt2011}
{Betancourt} M.,  {Stein} L.~C.,  2011, \mn@doi [arXiv e-prints]
  {10.48550/arXiv.1112.4118}, \href
  {https://ui.adsabs.harvard.edu/abs/2011arXiv1112.4118B} {p. arXiv:1112.4118}

\bibitem[\protect\citeauthoryear{Bingham et~al.,}{Bingham
  et~al.}{2019}]{Bingham2019}
Bingham E.,  et~al., 2019, The Journal of Machine Learning Research, 20, 973

\bibitem[\protect\citeauthoryear{Birrer et~al.,}{Birrer
  et~al.}{2021}]{Birrer2021}
Birrer S.,  et~al., 2021, \mn@doi [Journal of Open Source Software]
  {10.21105/joss.03283}, 6, 3283

\bibitem[\protect\citeauthoryear{{Bradley} et~al.,}{{Bradley}
  et~al.}{2020}]{photutils}
{Bradley} L.,  et~al., 2020, {astropy/photutils: 1.0.0},
  \mn@doi{10.5281/zenodo.4044744}

\bibitem[\protect\citeauthoryear{{Brennan} \& {Fraser}}{{Brennan} \&
  {Fraser}}{2022}]{Brennan2022}
{Brennan} S.~J.,  {Fraser} M.,  2022, \mn@doi [\aap]
  {10.1051/0004-6361/202243067}, \href
  {https://ui.adsabs.harvard.edu/abs/2022A&A...667A..62B} {667, A62}

\bibitem[\protect\citeauthoryear{Burger \& Burge}{Burger \&
  Burge}{2010}]{Burger2010}
Burger W.,  Burge M.,  2010, Principles of Digital Image Processing: Core
  Algorithms.
Undergraduate Topics in Computer Science, Springer London, \url
  {https://books.google.ca/books?id=s5CBZLBakawC}

\bibitem[\protect\citeauthoryear{{Ciambur}}{{Ciambur}}{2015}]{Ciambur2015}
{Ciambur} B.~C.,  2015, \mn@doi [\apj] {10.1088/0004-637X/810/2/120}, \href
  {https://ui.adsabs.harvard.edu/abs/2015ApJ...810..120C} {810, 120}

\bibitem[\protect\citeauthoryear{{Courteau}}{{Courteau}}{1996}]{Courteau1996}
{Courteau} S.,  1996, \mn@doi [\apjs] {10.1086/192281}, \href
  {https://ui.adsabs.harvard.edu/#abs/1996ApJS..103..363C} {103, 363}

\bibitem[\protect\citeauthoryear{{Dey} et~al.,}{{Dey} et~al.}{2019}]{Dey2019}
{Dey} A.,  et~al., 2019, \mn@doi [\aj] {10.3847/1538-3881/ab089d}, \href
  {https://ui.adsabs.harvard.edu/abs/2019AJ....157..168D} {157, 168}

\bibitem[\protect\citeauthoryear{{Ding} et~al.,}{{Ding}
  et~al.}{2020}]{Ding2020}
{Ding} X.,  et~al., 2020, \mn@doi [\apj] {10.3847/1538-4357/ab5b90}, \href
  {https://ui.adsabs.harvard.edu/abs/2020ApJ...888...37D} {888, 37}

\bibitem[\protect\citeauthoryear{Duane, Kennedy, Pendleton  \& Roweth}{Duane
  et~al.}{1987}]{Duane1987}
Duane S.,  Kennedy A.,  Pendleton B.~J.,   Roweth D.,  1987, \mn@doi [Physics
  Letters B] {https://doi.org/10.1016/0370-2693(87)91197-X}, 195, 216

\bibitem[\protect\citeauthoryear{{Dubath} et~al.,}{{Dubath}
  et~al.}{2017}]{EUCLID2017}
{Dubath} P.,  et~al., 2017, in {Brescia} M.,  {Djorgovski} S.~G.,  {Feigelson}
  E.~D.,  {Longo} G.,   {Cavuoti} S.,  eds,  IAU Symposium Vol. 325,
  Astroinformatics. pp 73--82 (\mn@eprint {arXiv} {1701.08158}),
  \mn@doi{10.1017/S1743921317001521}

\bibitem[\protect\citeauthoryear{{Erwin}}{{Erwin}}{2015}]{Erwin2015}
{Erwin} P.,  2015, \mn@doi [\apj] {10.1088/0004-637X/799/2/226}, \href
  {https://ui.adsabs.harvard.edu/abs/2015ApJ...799..226E} {799, 226}

\bibitem[\protect\citeauthoryear{Gavin}{Gavin}{2022}]{Gavin2022}
Gavin H.~P.,  2022, Department of Civil and Environmental Engineering, Duke
  University Durham, NC USA

\bibitem[\protect\citeauthoryear{{Gilmozzi}}{{Gilmozzi}}{2005}]{Gilmozzi2005}
{Gilmozzi} R.,  2005, \mn@doi [Experimental Astronomy]
  {10.1007/s10686-005-9007-0}, \href
  {https://ui.adsabs.harvard.edu/abs/2005ExA....19....5G} {19, 5}

\bibitem[\protect\citeauthoryear{{Gunes Baydin}, {Pearlmutter}, {Andreyevich
  Radul}  \& {Siskind}}{{Gunes Baydin} et~al.}{2015}]{Gunes2015}
{Gunes Baydin} A.,  {Pearlmutter} B.~A.,  {Andreyevich Radul} A.,   {Siskind}
  J.~M.,  2015, \mn@doi [arXiv e-prints] {10.48550/arXiv.1502.05767}, \href
  {https://ui.adsabs.harvard.edu/abs/2015arXiv150205767G} {p. arXiv:1502.05767}

\bibitem[\protect\citeauthoryear{Harris et~al.,}{Harris
  et~al.}{2020}]{harris2020numpy}
Harris C.~R.,  et~al., 2020, \mn@doi [Nature] {10.1038/s41586-020-2649-2}, 585,
  357

\bibitem[\protect\citeauthoryear{{H{\"a}u{\ss}ler} et~al.,}{{H{\"a}u{\ss}ler}
  et~al.}{2013}]{Hausler2013}
{H{\"a}u{\ss}ler} B.,  et~al., 2013, \mn@doi [\mnras] {10.1093/mnras/sts633},
  \href {https://ui.adsabs.harvard.edu/abs/2013MNRAS.430..330H} {430, 330}

\bibitem[\protect\citeauthoryear{{Hoffman} \& {Gelman}}{{Hoffman} \&
  {Gelman}}{2011}]{Hoffman2011}
{Hoffman} M.~D.,  {Gelman} A.,  2011, \mn@doi [arXiv e-prints]
  {10.48550/arXiv.1111.4246}, \href
  {https://ui.adsabs.harvard.edu/abs/2011arXiv1111.4246H} {p. arXiv:1111.4246}

\bibitem[\protect\citeauthoryear{{Infante-Sainz}, {Trujillo}  \&
  {Rom{\'a}n}}{{Infante-Sainz} et~al.}{2020}]{InfanteSainz2020}
{Infante-Sainz} R.,  {Trujillo} I.,   {Rom{\'a}n} J.,  2020, \mn@doi [\mnras]
  {10.1093/mnras/stz3111}, \href
  {https://ui.adsabs.harvard.edu/abs/2020MNRAS.491.5317I} {491, 5317}

\bibitem[\protect\citeauthoryear{{Ivezi{\'c}} et~al.,}{{Ivezi{\'c}}
  et~al.}{2019}]{Ivezic2019}
{Ivezi{\'c}} {\v Z}.,  et~al., 2019, \mn@doi [\apj] {10.3847/1538-4357/ab042c},
  \href {http://adsabs.harvard.edu/abs/2019ApJ...873..111I} {873, 111}

\bibitem[\protect\citeauthoryear{{Jarvis} \& {Jain}}{{Jarvis} \&
  {Jain}}{2004}]{Jarvis2004}
{Jarvis} M.,  {Jain} B.,  2004, \mn@doi [arXiv e-prints]
  {10.48550/arXiv.astro-ph/0412234}, \href
  {https://ui.adsabs.harvard.edu/abs/2004astro.ph.12234J} {pp
  astro--ph/0412234}

\bibitem[\protect\citeauthoryear{{Jia}, {Wu}, {Yi}, {Cai}  \& {Cai}}{{Jia}
  et~al.}{2020}]{Jia2020}
{Jia} P.,  {Wu} X.,  {Yi} H.,  {Cai} B.,   {Cai} D.,  2020, \mn@doi [\aj]
  {10.3847/1538-3881/ab7b79}, \href
  {https://ui.adsabs.harvard.edu/abs/2020AJ....159..183J} {159, 183}

\bibitem[\protect\citeauthoryear{{Kingma} \& {Ba}}{{Kingma} \&
  {Ba}}{2014}]{Kingma2014}
{Kingma} D.~P.,  {Ba} J.,  2014, \mn@doi [arXiv e-prints]
  {10.48550/arXiv.1412.6980}, \href
  {https://ui.adsabs.harvard.edu/abs/2014arXiv1412.6980K} {p. arXiv:1412.6980}

\bibitem[\protect\citeauthoryear{{Krist}}{{Krist}}{1993}]{Krist1993}
{Krist} J.,  1993, in {Hanisch} R.~J.,  {Brissenden} R.~J.~V.,   {Barnes} J.,
  eds,  Astronomical Society of the Pacific Conference Series Vol. 52,
  Astronomical Data Analysis Software and Systems II. p.~536

\bibitem[\protect\citeauthoryear{{Lang}}{{Lang}}{2014}]{Lang2014}
{Lang} D.,  2014, \mn@doi [\aj] {10.1088/0004-6256/147/5/108}, \href
  {https://ui.adsabs.harvard.edu/abs/2014AJ....147..108L} {147, 108}

\bibitem[\protect\citeauthoryear{{Lang}, {Hogg}  \& {Mykytyn}}{{Lang}
  et~al.}{2016}]{Lang2016}
{Lang} D.,  {Hogg} D.~W.,   {Mykytyn} D.,  2016, {The Tractor: Probabilistic
  astronomical source detection and measurement}, Astrophysics Source Code
  Library, record ascl:1604.008 (\mn@eprint {ascl} {1604.008})

\bibitem[\protect\citeauthoryear{{Lauer}}{{Lauer}}{1985}]{Lauer1985}
{Lauer} T.~R.,  1985, \mn@doi [\apjs] {10.1086/191011}, \href
  {https://ui.adsabs.harvard.edu/abs/1985ApJS...57..473L} {57, 473}

\bibitem[\protect\citeauthoryear{{Lauer} et~al.,}{{Lauer}
  et~al.}{1995}]{Lauer1995}
{Lauer} T.~R.,  et~al., 1995, \mn@doi [\aj] {10.1086/117719}, \href
  {https://ui.adsabs.harvard.edu/abs/1995AJ....110.2622L} {110, 2622}

\bibitem[\protect\citeauthoryear{{Liaudat}, {Starck}, {Kilbinger}  \&
  {Frugier}}{{Liaudat} et~al.}{2023}]{Liaudat2023}
{Liaudat} T.,  {Starck} J.-L.,  {Kilbinger} M.,   {Frugier} P.-A.,  2023,
  \mn@doi [arXiv e-prints] {10.48550/arXiv.2306.07996}, \href
  {https://ui.adsabs.harvard.edu/abs/2023arXiv230607996L} {p. arXiv:2306.07996}

\bibitem[\protect\citeauthoryear{{Lu}, {Zhang}, {Dong}, {Li}, {Liu}, {Fu}, {Li}
   \& {Fan}}{{Lu} et~al.}{2017}]{Lu2017}
{Lu} T.,  {Zhang} J.,  {Dong} F.,  {Li} Y.,  {Liu} D.,  {Fu} L.,  {Li} G.,
  {Fan} Z.,  2017, \mn@doi [\aj] {10.3847/1538-3881/aa661e}, \href
  {https://ui.adsabs.harvard.edu/abs/2017AJ....153..197L} {153, 197}

\bibitem[\protect\citeauthoryear{{Marmo} \& {Bertin}}{{Marmo} \&
  {Bertin}}{2008}]{Marmo2008}
{Marmo} C.,  {Bertin} E.,  2008, in {Argyle} R.~W.,  {Bunclark} P.~S.,
  {Lewis} J.~R.,  eds,  Astronomical Society of the Pacific Conference Series
  Vol. 394, Astronomical Data Analysis Software and Systems XVII. p.~619

\bibitem[\protect\citeauthoryear{Marquardt}{Marquardt}{1963}]{Marquardt1963}
Marquardt D.~W.,  1963, \mn@doi [Journal of the Society for Industrial and
  Applied Mathematics] {10.1137/0111030}, 11, 431

\bibitem[\protect\citeauthoryear{{Mihos}}{{Mihos}}{2019}]{Mihos2019}
{Mihos} J.~C.,  2019, \mn@doi [arXiv e-prints] {10.48550/arXiv.1909.09456},
  \href {https://ui.adsabs.harvard.edu/abs/2019arXiv190909456M} {p.
  arXiv:1909.09456}

\bibitem[\protect\citeauthoryear{{Moffat}}{{Moffat}}{1969}]{Moffat1969}
{Moffat} A.~F.~J.,  1969, \aap, \href
  {https://ui.adsabs.harvard.edu/abs/1969A&A.....3..455M} {3, 455}

\bibitem[\protect\citeauthoryear{{Morrissey} et~al.,}{{Morrissey}
  et~al.}{2007}]{Morrissey2007}
{Morrissey} P.,  et~al., 2007, \mn@doi [\apjs] {10.1086/520512}, \href
  {https://ui.adsabs.harvard.edu/abs/2007ApJS..173..682M} {173, 682}

\bibitem[\protect\citeauthoryear{{M{\"u}ller-Bravo} \&
  {Galbany}}{{M{\"u}ller-Bravo} \& {Galbany}}{2022}]{Muller2022}
{M{\"u}ller-Bravo} T.,  {Galbany} L.,  2022, \mn@doi [The Journal of Open
  Source Software] {10.21105/joss.04508}, \href
  {https://ui.adsabs.harvard.edu/abs/2022JOSS....7.4508M} {7, 4508}

\bibitem[\protect\citeauthoryear{Nightingale et~al.,}{Nightingale
  et~al.}{2023}]{Nightingale2023}
Nightingale J.~W.,  et~al., 2023, \mn@doi [Journal of Open Source Software]
  {10.21105/joss.04475}, 8, 4475

\bibitem[\protect\citeauthoryear{{Nikolic}}{{Nikolic}}{2018}]{Nikolic2018}
{Nikolic} B.,  2018, \mn@doi [arXiv e-prints] {10.48550/arXiv.1805.07439},
  \href {https://ui.adsabs.harvard.edu/abs/2018arXiv180507439N} {p.
  arXiv:1805.07439}

\bibitem[\protect\citeauthoryear{{Pasha} \& {Miller}}{{Pasha} \&
  {Miller}}{2023}]{Pasha2023}
{Pasha} I.,  {Miller} T.~B.,  2023, \mn@doi [arXiv e-prints]
  {10.48550/arXiv.2306.05454}, \href
  {https://ui.adsabs.harvard.edu/abs/2023arXiv230605454P} {p. arXiv:2306.05454}

\bibitem[\protect\citeauthoryear{{Pence}, {Chiappetti}, {Page}, {Shaw}  \&
  {Stobie}}{{Pence} et~al.}{2010}]{Pence2010}
{Pence} W.~D.,  {Chiappetti} L.,  {Page} C.~G.,  {Shaw} R.~A.,   {Stobie} E.,
  2010, \mn@doi [\aap] {10.1051/0004-6361/201015362}, \href
  {https://ui.adsabs.harvard.edu/abs/2010A&A...524A..42P} {524, A42}

\bibitem[\protect\citeauthoryear{{Peng}, {Ho}, {Impey}  \& {Rix}}{{Peng}
  et~al.}{2010}]{Peng2010}
{Peng} C.~Y.,  {Ho} L.~C.,  {Impey} C.~D.,   {Rix} H.-W.,  2010, \mn@doi [\aj]
  {10.1088/0004-6256/139/6/2097}, \href
  {https://ui.adsabs.harvard.edu/abs/2010AJ....139.2097P} {139, 2097}

\bibitem[\protect\citeauthoryear{{Perrin}, {Sivaramakrishnan}, {Lajoie},
  {Elliott}, {Pueyo}, {Ravindranath}  \& {Albert}}{{Perrin}
  et~al.}{2014}]{Perrin2014}
{Perrin} M.~D.,  {Sivaramakrishnan} A.,  {Lajoie} C.-P.,  {Elliott} E.,
  {Pueyo} L.,  {Ravindranath} S.,   {Albert} L.,  2014, in {Oschmann}
  Jacobus~M. J.,  {Clampin} M.,  {Fazio} G.~G.,   {MacEwen} H.~A.,  eds,
  Society of Photo-Optical Instrumentation Engineers (SPIE) Conference Series
  Vol. 9143, Space Telescopes and Instrumentation 2014: Optical, Infrared, and
  Millimeter Wave. p. 91433X, \mn@doi{10.1117/12.2056689}

\bibitem[\protect\citeauthoryear{Petersen, Pedersen  et~al.}{Petersen
  et~al.}{2008}]{petersen2008matrix}
Petersen K.~B.,  Pedersen M.~S.,   et~al., 2008, Technical University of
  Denmark, 7, 510

\bibitem[\protect\citeauthoryear{{Remy}, {Lanusse}, {Jeffrey}, {Liu}, {Starck},
  {Osato}  \& {Schrabback}}{{Remy} et~al.}{2022}]{Remy2022}
{Remy} B.,  {Lanusse} F.,  {Jeffrey} N.,  {Liu} J.,  {Starck} J.-L.,  {Osato}
  K.,   {Schrabback} T.,  2022, \mn@doi [arXiv e-prints]
  {10.48550/arXiv.2201.05561}, \href
  {https://ui.adsabs.harvard.edu/abs/2022arXiv220105561R} {p. arXiv:2201.05561}

\bibitem[\protect\citeauthoryear{{Rigamonti}, {Dotti}, {Covino}, {Haardt},
  {Cortese}, {Landoni}  \& {Varisco}}{{Rigamonti} et~al.}{2023}]{Rigamonti2023}
{Rigamonti} F.,  {Dotti} M.,  {Covino} S.,  {Haardt} F.,  {Cortese} L.,
  {Landoni} M.,   {Varisco} L.,  2023, \mn@doi [arXiv e-prints]
  {10.48550/arXiv.2305.03762}, \href
  {https://ui.adsabs.harvard.edu/abs/2023arXiv230503762R} {p. arXiv:2305.03762}

\bibitem[\protect\citeauthoryear{{Roberts} \& {Rosenthal}}{{Roberts} \&
  {Rosenthal}}{2014}]{Roberts2014}
{Roberts} G.~O.,  {Rosenthal} J.~S.,  2014, \mn@doi [arXiv e-prints]
  {10.48550/arXiv.1411.0712}, \href
  {https://ui.adsabs.harvard.edu/abs/2014arXiv1411.0712R} {p. arXiv:1411.0712}

\bibitem[\protect\citeauthoryear{{Robotham}, {Taranu}, {Tobar}, {Moffett}  \&
  {Driver}}{{Robotham} et~al.}{2017}]{Robotham2017}
{Robotham} A.~S.~G.,  {Taranu} D.~S.,  {Tobar} R.,  {Moffett} A.,   {Driver}
  S.~P.,  2017, \mn@doi [\mnras] {10.1093/mnras/stw3039}, \href
  {https://ui.adsabs.harvard.edu/abs/2017MNRAS.466.1513R} {466, 1513}

\bibitem[\protect\citeauthoryear{{Robotham}, {Davies}, {Driver}, {Koushan},
  {Taranu}, {Casura}  \& {Liske}}{{Robotham} et~al.}{2018}]{Robotham2018}
{Robotham} A.~S.~G.,  {Davies} L.~J.~M.,  {Driver} S.~P.,  {Koushan} S.,
  {Taranu} D.~S.,  {Casura} S.,   {Liske} J.,  2018, \mn@doi [\mnras]
  {10.1093/mnras/sty440}, \href
  {https://ui.adsabs.harvard.edu/abs/2018MNRAS.476.3137R} {476, 3137}

\bibitem[\protect\citeauthoryear{{Robotham}, {Bellstedt}  \&
  {Driver}}{{Robotham} et~al.}{2022}]{Robotham2022}
{Robotham} A.~S.~G.,  {Bellstedt} S.,   {Driver} S.~P.,  2022, \mn@doi [\mnras]
  {10.1093/mnras/stac1032}, \href
  {https://ui.adsabs.harvard.edu/abs/2022MNRAS.513.2985R} {513, 2985}

\bibitem[\protect\citeauthoryear{{Rowe} et~al.,}{{Rowe}
  et~al.}{2015}]{Rowe2015}
{Rowe} B.~T.~P.,  et~al., 2015, \mn@doi [Astronomy and Computing]
  {10.1016/j.ascom.2015.02.002}, \href
  {https://ui.adsabs.harvard.edu/abs/2015A&C....10..121R} {10, 121}

\bibitem[\protect\citeauthoryear{{Schmitz} et~al.,}{{Schmitz}
  et~al.}{2020}]{Schmitz2020}
{Schmitz} M.~A.,  et~al., 2020, \mn@doi [\aap] {10.1051/0004-6361/201936094},
  \href {https://ui.adsabs.harvard.edu/abs/2020A&A...636A..78S} {636, A78}

\bibitem[\protect\citeauthoryear{{Sellwood} \& {Spekkens}}{{Sellwood} \&
  {Spekkens}}{2015}]{Sellwood2015}
{Sellwood} J.~A.,  {Spekkens} K.,  2015, \mn@doi [arXiv e-prints]
  {10.48550/arXiv.1509.07120}, \href
  {https://ui.adsabs.harvard.edu/abs/2015arXiv150907120S} {p. arXiv:1509.07120}

\bibitem[\protect\citeauthoryear{{S{\'e}rsic}}{{S{\'e}rsic}}{1963}]{sersic1963}
{S{\'e}rsic} J.~L.,  1963, Boletin de la Asociacion Argentina de Astronomia La
  Plata Argentina, \href
  {https://ui.adsabs.harvard.edu/abs/1963BAAA....6...41S} {6, 41}

\bibitem[\protect\citeauthoryear{Shannon}{Shannon}{1949}]{Shannon1949}
Shannon C.,  1949, \mn@doi [Proceedings of the IRE]
  {10.1109/JRPROC.1949.232969}, 37, 10

\bibitem[\protect\citeauthoryear{{Simard} et~al.,}{{Simard}
  et~al.}{2002}]{Simard2002}
{Simard} L.,  et~al., 2002, \mn@doi [\apjs] {10.1086/341399}, \href
  {https://ui.adsabs.harvard.edu/abs/2002ApJS..142....1S} {142, 1}

\bibitem[\protect\citeauthoryear{{Smith}, {Arora}, {Stone}, {Courteau}  \&
  {Geach}}{{Smith} et~al.}{2021}]{Smith2021}
{Smith} M.~J.,  {Arora} N.,  {Stone} C.,  {Courteau} S.,   {Geach} J.~E.,
  2021, \mn@doi [\mnras] {10.1093/mnras/stab424}, \href
  {https://ui.adsabs.harvard.edu/abs/2021MNRAS.503...96S} {503, 96}

\bibitem[\protect\citeauthoryear{{Smith}, {Geach}, {Jackson}, {Arora}, {Stone}
  \& {Courteau}}{{Smith} et~al.}{2022}]{Smith2022}
{Smith} M.~J.,  {Geach} J.~E.,  {Jackson} R.~A.,  {Arora} N.,  {Stone} C.,
  {Courteau} S.,  2022, \mn@doi [\mnras] {10.1093/mnras/stac130}, \href
  {https://ui.adsabs.harvard.edu/abs/2022MNRAS.511.1808S} {511, 1808}

\bibitem[\protect\citeauthoryear{{Stone}, {Arora}, {Courteau}  \&
  {Cuillandre}}{{Stone} et~al.}{2021}]{Stone2021b}
{Stone} C.~J.,  {Arora} N.,  {Courteau} S.,   {Cuillandre} J.-C.,  2021,
  \mn@doi [\mnras] {10.1093/mnras/stab2709}, \href
  {https://ui.adsabs.harvard.edu/abs/2021MNRAS.508.1870S} {508, 1870}

\bibitem[\protect\citeauthoryear{{Sun}, {Bohm Agostini}, {Dong}  \&
  {Kaeli}}{{Sun} et~al.}{2019}]{Sun2019}
{Sun} Y.,  {Bohm Agostini} N.,  {Dong} S.,   {Kaeli} D.,  2019, \mn@doi [arXiv
  e-prints] {10.48550/arXiv.1911.11313}, \href
  {https://ui.adsabs.harvard.edu/abs/2019arXiv191111313S} {p. arXiv:1911.11313}

\bibitem[\protect\citeauthoryear{Teeninga, Moschini, Trager  \&
  Wilkinson}{Teeninga et~al.}{2015}]{teeninga2015}
Teeninga P.,  Moschini U.,  Trager S.~C.,   Wilkinson M.~H.,  2015, in
  International Symposium on Mathematical Morphology and Its Applications to
  Signal and Image Processing. pp 157--168

\bibitem[\protect\citeauthoryear{{Tody}}{{Tody}}{1986}]{Tody1986}
{Tody} D.,  1986, in {Crawford} D.~L.,  ed.,  Society of Photo-Optical
  Instrumentation Engineers (SPIE) Conference Series Vol. 627, Instrumentation
  in astronomy VI. p.~733, \mn@doi{10.1117/12.968154}

\bibitem[\protect\citeauthoryear{{Tortorelli} \& {Mercurio}}{{Tortorelli} \&
  {Mercurio}}{2023}]{Tortorelli2023}
{Tortorelli} L.,  {Mercurio} A.,  2023, \mn@doi [arXiv e-prints]
  {10.48550/arXiv.2302.07890}, \href
  {https://ui.adsabs.harvard.edu/abs/2023arXiv230207890T} {p. arXiv:2302.07890}

\bibitem[\protect\citeauthoryear{{Transtrum} \& {Sethna}}{{Transtrum} \&
  {Sethna}}{2012}]{Transtrum2012}
{Transtrum} M.~K.,  {Sethna} J.~P.,  2012, \mn@doi [arXiv e-prints]
  {10.48550/arXiv.1201.5885}, \href
  {https://ui.adsabs.harvard.edu/abs/2012arXiv1201.5885T} {p. arXiv:1201.5885}

\bibitem[\protect\citeauthoryear{{Vikram}, {Wadadekar}, {Kembhavi}  \&
  {Vijayagovindan}}{{Vikram} et~al.}{2010}]{Vikram2010}
{Vikram} V.,  {Wadadekar} Y.,  {Kembhavi} A.~K.,   {Vijayagovindan} G.~V.,
  2010, \mn@doi [\mnras] {10.1111/j.1365-2966.2010.17426.x}, \href
  {https://ui.adsabs.harvard.edu/abs/2010MNRAS.409.1379V} {409, 1379}

\bibitem[\protect\citeauthoryear{{Walmsley} et~al.,}{{Walmsley}
  et~al.}{2022}]{Walmsley2022}
{Walmsley} M.,  et~al., 2022, \mn@doi [\mnras] {10.1093/mnras/stab2093}, \href
  {https://ui.adsabs.harvard.edu/abs/2022MNRAS.509.3966W} {509, 3966}

\bibitem[\protect\citeauthoryear{{Willmer}}{{Willmer}}{2018}]{Wilmer2018}
{Willmer} C. N.~A.,  2018, \mn@doi [\apjs] {10.3847/1538-4365/aabfdf}, \href
  {https://ui.adsabs.harvard.edu/abs/2018ApJS..236...47W} {236, 47}

\makeatother
\end{thebibliography}

\appendix

\section{SED from multiband fit}
\label{app:sed}

\begin{figure}
    \centering
    \includegraphics[width = \columnwidth]{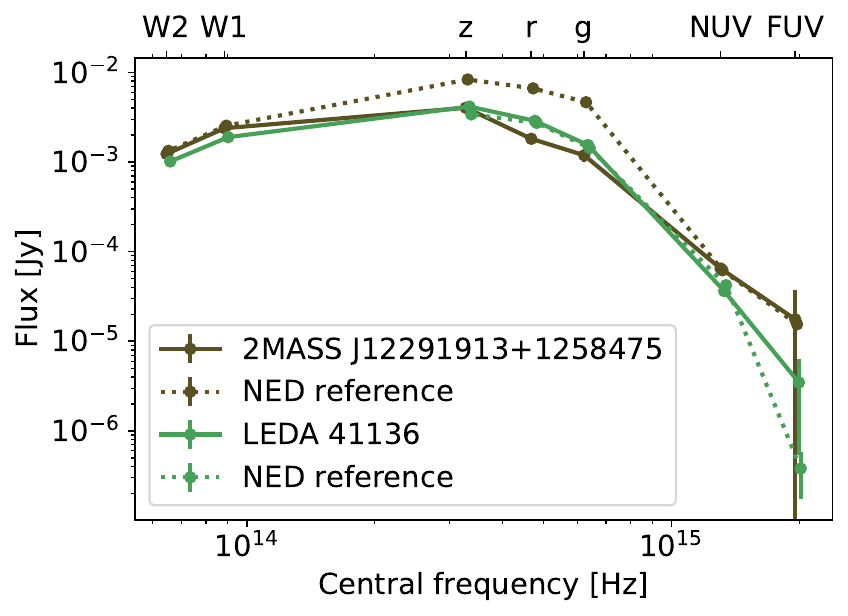}
    \caption{SED for LEDA41136 and a nearby star (\emph{2MASS J12291913+1258475}). 
    The solid lines show SEDs from \AP, while the dashed lines show SEDs extracted from the literature via NED. 
    Fluxes are in Janskys and the central frequency for each bandpass is quoted in Hertz. 
    Each bandpass is offset horizontally by a small amount to reduce crowding.}
    \label{fig:sed}
\end{figure}

\Fig{sed} presents the extracted spectral energy distribution (SED) from the fit in \Fig{multibandfit}.
The results are also compared to fluxes available in NED\footnote{The NASA/IPAC Extragalactic Database (NED) is funded by the National Aeronautics and Space Administration and operated by the California Institute of Technology.}, and all are converted to Janskys~\citep{Wilmer2018}.
Overall the agreement between \AP and the literature is remarkable.
Notably, the fluxes for the star \emph{2MASS J12291913+1258475} in the g-, r-, and z-bands are significantly offset, though this is not surprising given that most of the star is saturated in these bands (and hence masked).
All other fluxes are consistent with each other up to a reasonable uncertainty limit.

\section{Model Organization Chart}

\begin{figure*}
    \centering
    \includegraphics[width=0.9\textwidth]{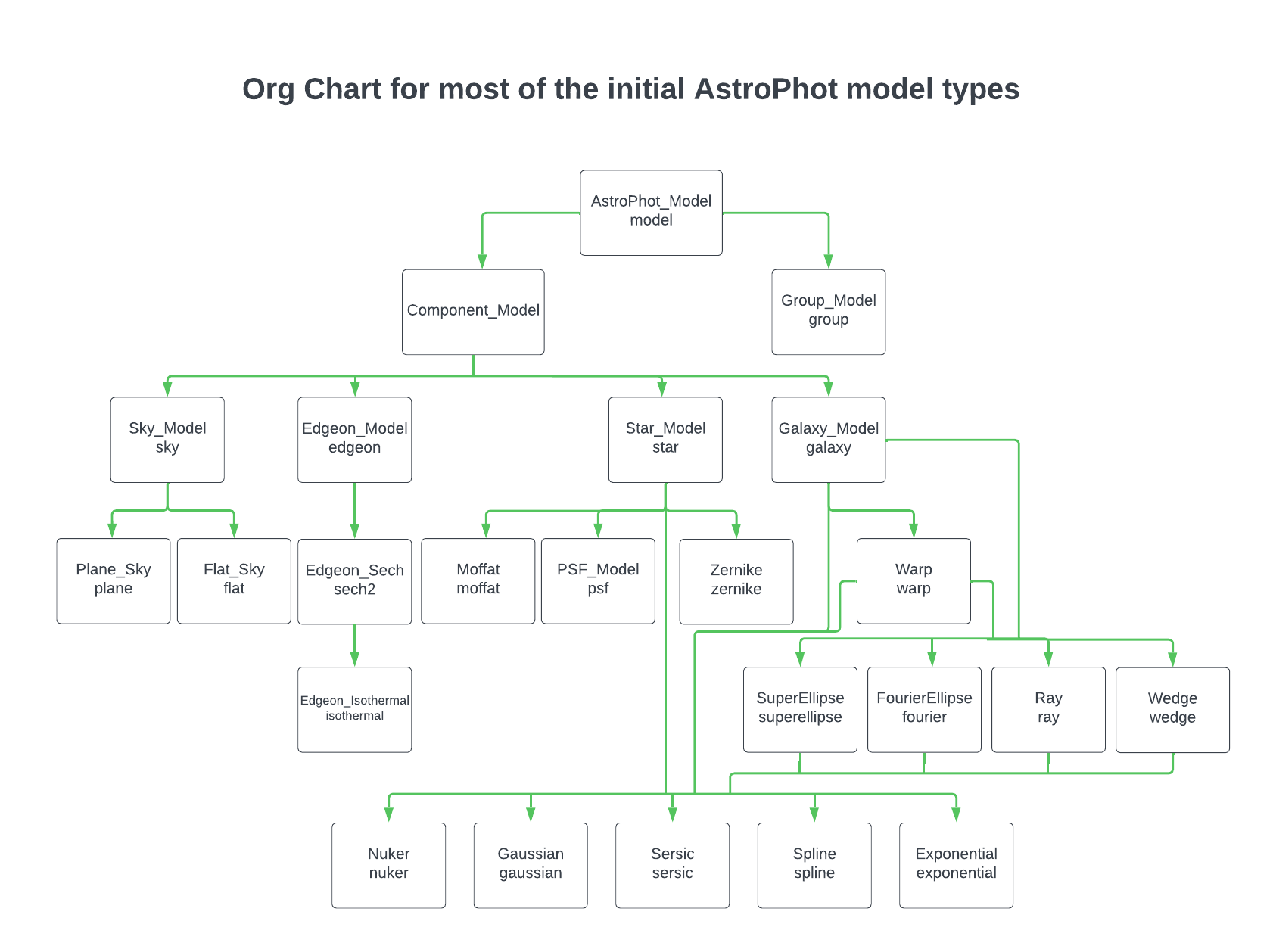}
    \caption{Organization chart of available models in \AP (expected to grow over time). 
    The flow chart shows the inheritance structure of the models with their naming convention. 
    In each box, the upper name is the Python object for that model, the lower name is used to construct the string representation of the model.
    One can recover the full name of a model by following any path backwards through the flowchart and collecting the lower names.  
    A simple example is ``group~model'' while a more complex example would be ``sersic~ray~warp~galaxy~model''. 
    One can request from \AP a list of all currently available models.
    }
    \label{fig:modelorgchart}
\end{figure*}

The object-oriented framework inherent to \AP models enables users to branch off and construct new custom models.
\Fig{modelorgchart} presents this inheritance structure as a flow chart.
Starting from any location in the flowchart, a new model can be created with its associated properties plus any user defined features.
For example one could start from the ``Galaxy\_Model'' object and add a new radial SB profile.
All transformations related to fitting the center, q, and PA would already be handled leaving only the SB profile parameters for the user to define.
Examples of constructing custom models in \AP are given in the code documentation.
\Fig{modelzoo} shows examples of currently available models in \AP. 

\section{Example spline configuration file}

Interfacing with the \AP image modelling package can take on two distinct forms. 
The primary method of operating \AP is within an object-oriented framework, where one instantiates model objects and then calls methods to perform initialization, fitting, sampling, etc.
Extensive documentation on this interface is available in the form of tutorial examples and more traditional documentation.
For simpler fitting tasks, one may also use a configuration file interface.
Configuration files can take advantage of standard features in \AP, though advanced operations such as joint models require more attention.

\Fig{configfile} gives an illustrative example of a configuration file designed for fitting a galaxy image similar to that presented in \Sec{examplespline}\footnote{To fully match \Sec{examplespline}, simply swap \emph{spline galaxy model} with \emph{spline warp galaxy model}. 
The execution time for the latter is greater since it is a more complex model.}.
This example serves as a practical guide to understand the input format in the configuration files.
The beginning of the configuration file downloads an example image from the DESI Legacy Imaging Survey~\citep{Dey2019}.
This step would not be needed if the data were already loaded on the computer.
Additionally, the file contains information regarding the image pixel scale and the magnitude zero-point system employed.

The next sections of the configuration file address the model components and their corresponding parameters. 
For each component, such as a \emph{spline galaxy model} or a \emph{sersic galaxy model}, the file outlines the sky fitting region (no specification means fit the whole image).
This is done by giving the pixel coordinates of a bounding box around the fitting region.
Initial parameter estimates and constraints, including the effective radius, axis ratio, position angle, and S{\'e}rsic index are automatically determined by \AP. 
These parameters can be manually adjusted to tailor the fitting process according to the specific scientific objectives and the characteristics of the galaxy image.
As an example, one of the sub models has all of its parameters manually initialized.

The configuration file can also include the optimization algorithm settings, such as convergence criteria, maximum iterations, and damping parameters. 
Here the level of output verbosity and maximum number of iterations allowed is specified.
The default optimization algorithm used by configuration file inputs is the LM method since it is the most generally applicable.

Finally, the output configuration is specified. 
Here the model parameters are to be saved to a file named \emph{spline\_model.yaml} where YAML is a common format for storing data which is a compromise of machine and human readable.
FITS file images of the model evaluated at the optimal parameters and a residual image are saved to separate files.
If no path for the model or residual image is specified, then they are not saved. 

\begin{figure}
    \centering
\begin{verbatim}
# Download the data
from astropy.io import fits
hdu = fits.open("https://www.legacysurvey.org/viewer/\
fits-cutout?ra=36.3684&dec=-25.6389&size=700&layer=\
ls-dr9&pixscale=0.262&bands=r")
hdu.writeto("ESO479-G1.fits", overwrite = True)

# Image features
ap_target_file = "ESO479-G1.fits"
ap_target_pixelscale = 0.262
ap_target_zeropoint = 22.5

# Main galaxy model
ap_model_ESO479_G1 = {"model_type": "spline galaxy model"}

# Interloper galaxies
ap_model_sub1 = {"model_type": "sersic galaxy model",
                 "window": [[480, 590],[555, 665]]}
ap_model_sub2 = {"model_type": "sersic galaxy model",
                 "window": [[572, 630],[534, 611]]}
ap_model_sub3 = {"model_type": "sersic galaxy model",
                 "window": [[183, 240],[0, 46]]}
ap_model_sub4 = {"model_type": "sersic galaxy model",
                 "window": [[103, 167], [557, 610]]}
ap_model_sub5 = {
    "model_type": "sersic galaxy model",
    "window": [[336, 385], [15, 65]],
    "parameters": {
        "center": [95.,10.], # arcsec from bottom corner
        "q": 0.9, # b / a
        "PA": 2.85, # radians
        "n": 1., # sersic index
        "Re": 0.6, # arcsec
        "Ie": 1.3, # log10(flux / arcsec^2)
    }}

# Optimizer
ap_optimizer_kwargs = {"verbose": 1, "max_iter": 100}

# Output
ap_saveto_model = "spline_model.yaml"
ap_saveto_model_image = "spline_model_image.fits"
ap_saveto_model_residual = "spline_model_residual.fits"
\end{verbatim}
    \caption{
    Example configuration file for reproduction of the results in \Sec{examplespline}. 
    This file should be saved as a Python file (ending in \emph{.py}) and initiated by calling \emph{astrophot config.py} on the file. 
    This is the easiest way to use \AP.
    For more flexibility, an object-oriented interface is available for which several tutorial notebooks and extensive documentation exist.}
    \label{fig:configfile}
\end{figure}

\bsp	
\label{lastpage}
\end{document}